# Advancements in Superconducting Microwave Cavities and Qubits for Quantum Information Systems

Alex Krasnok[1*], Pashupati Dhakal[2], Arkady Fedorov[3], Pedro Frigola[4], Michael Kelly[5], and Sergey Kutsaev[4]

[1] *Department of Electrical and Computer Engineering, Florida International University, Miami, FL 33174, USA*

[2]*Thomas Jefferson National Accelerator Facility, Newport News, VA 23606, USA*

[3]*ARC Centre for Engineered Quantum System, School of Mathematics and Physics, University of Queensland, Brisbane QLD 4072, Australia*

[4]*RadiaBeam LLC, Santa Monica, CA 90403, USA*

[5]*Argonne National Laboratory, Lemont, IL 60439, USA*

**To whom correspondence should be addressed:* akrasnok@fiu.edu

**Abstract**

Superconducting microwave cavities with ultra-high Q-factors are revolutionizing the field of quantum computing, offering long coherence times exceeding 1 ms, which is critical for realizing scalable multi-qubit quantum systems with low error rates. In this work, we provide an in-depth analysis of recent advances in ultra-high Q-factor cavities, integration of Josephson junction-based qubits, and bosonic-encoded qubits in 3D cavities. We carefully examine the sources of quantum state dephasing caused by damping and noise mechanisms in cavities and qubits, highlighting the critical challenges that need to be addressed to achieve even higher coherence times. We critically survey the latest progress made in implementing single 3D qubits using superconducting materials, normal metals, and multi-qubit and multi-state quantum systems. Our work sheds light on the promising future of this research area, including novel materials for cavities and qubits, modes with nontrivial topological properties, error correction techniques for bosonic qubits, and new light-matter interaction effects.

## 1. Introduction

The advent of quantum computers (QCs) has sparked a revolution in computing with their demonstrated ability to solve problems that are beyond the capability of classical electronic computers because of quantum coherence and parallel computation[1–8]. This emerging field has opened up a wide range of potential applications, including first-principle calculations of atomic quantum spectra and materials design, security, and data search[3,9–11]. The computing capacity of QCs is expected to exceed that of classical computers, which makes them promising for solving previously unfeasible problems efficiently[6,12–14]. As quantum processors advance from the scale of 50-100 qubits to hundreds or more, a host of formidable challenges emerge, including issues related to quantum error correction (QEC), qubit connectivity, and hardware scalability. Foremost among these challenges is the urgent need for qubits with long coherence times to enable quantum error correction and facilitate the execution of large-scale quantum circuits. However, the inherent fragility of quantum states renders them vulnerable to destruction by even the slightest environmental interference, posing a major hurdle to achieving universal quantum computation[15]. The development of a practical quantum computer that can support large circuit depth requires the implementation of logical qubits shielded by QEC against unwanted or uncontrolled errors[16,17]. Such an endeavor is expected to demand the lion's share of computational resources for error correction. Consequently, the creation of a logical qubit with a coherence time longer than its individual physical components is one of the most pressing and challenging objectives in the field of quantum information processing[18].

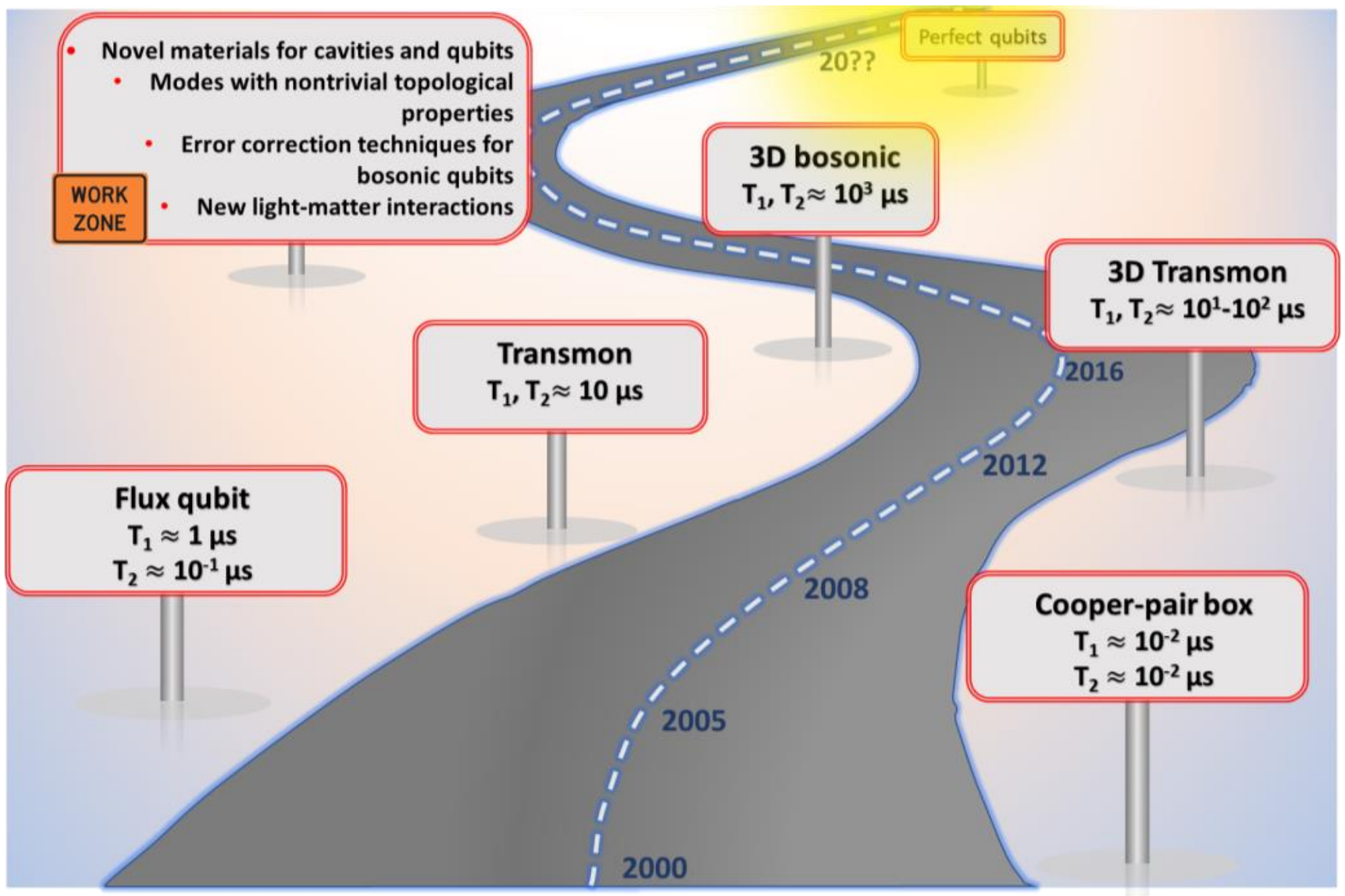

**Figure 1**. **Pathway to flawless qubits.** Advancements in $T_1$ and $T_2$ lifetimes of superconducting qubits in the last two decades and perspective ways of their improvement: novel materials for cavities and qubits, modes with nontrivial topological properties, error correction techniques for bosonic qubits and new light-matter interaction effects (anomalies in EM scattering, virtual excitation).

Although there are other natural candidates for quantum computing, such as trapped ions[19,20], cold atoms[21–23], and NV centers[24], and quantum dots (QD)[25], superconducting qubits are considered as one of the most promising platforms. Superconducting qubits can be broadly categorized into two types: qubits encoded in the supercurrents of circuits with *Josephson junctions* (JJ) and *bosonic-encoded qubits*, where the quantum information is encoded in the superposition of bosonic states. JJ-qubits, also known as artificial quantum systems, possess a macroscopic size ranging 50-500 μm. These qubits are made up of superconducting circuits that contain JJs, which are essential non-linear electrical components allowing the tunneling of superconducting Cooper pairs. Additionally, these circuits comprise inductor and capacitor components. These components are carefully patterned on a substrate, which is usually composed of silicon or sapphire, ensuring optimal performance.

One of the major challenges facing superconducting quantum computing is its relatively short coherence time, that described by $T_1$ and $T_2$, which are the longitudinal and transverse relaxation times, respectively. These parameters characterize the time it takes for the quantum system to lose information due to environmental factors such as temperature, magnetic fields, and electrical noise, ultimately limiting the performance and scalability of quantum computers. There has been a relentless pursuit to improve the coherence time of superconducting qubits since their inception at the end of the last century[26], **Fig. 1**. The first devices possessed a very short quantum coherence lifetime of $10^{-2}$ μs. Since then, many groups worldwide conceived and implemented a variety of JJ qubits by varying with electrical circuits, e.g., by adding loops interrupted by one or more JJs or by adding capacitors. Research involving these variants helped the community shed light on what limits coherence times. Progress towards sustaining quantum effects in cQED circuits of growing complexity has been remarkable, with superconducting quantum processors rapidly growing in qubit count and functionality and demonstrating a five- orders-of-magnitude improvement[27,28], **Fig. 1**.

Although state-of-the-art 2D JJ-qubits typically have coherence times ranging from $10^1$-$10^2$ μs[28,29], this is inadequate for large-scale circuit execution. In contrast, standalone 3D superconducting cavities have demonstrated coherence times of ~$10^3$ μs and beyond, offering the potential for scalable multi-qubit quantum systems with long coherence and low error rates[30,31]. By combining high-Q 3D cavities with JJ qubits, the potential for multi-qubit quantum systems with long quantum coherence and low error rates seem enormous at the time, **Fig.1**. This approach presented a promising solution to the challenge of achieving long coherence times and improving error rates in quantum systems and showed the path to increase the lifetime for 2D architecture. This led to development 2D transmon qubits with coherence times of ~$10^2$ μs, that currently used in the most of existing small scale quantum computers.

Later the advantage in debeloping better qubits was primiraly associated with developing qubits of alternative topologies, such as fluxonium with coherence times of ~400 μs [32] and even ~1.5 ms[33], as well as implementing bosonic-encoded qubits (BEC), which utilize various encoding techniques. BECs primarily consist of a high-Q 3D resonator that is coupled through a non-linear *ancilla*, often another JJ-qubit. The primary purpose of the ancilla is to provide effective control and tomography of the encoded information within the BEC. This method enables the precise manipulation and measurement of the quantum state, further solidifying the potential of superconducting qubits in quantum computing applications. Furthermore, the development of error-corrected qubits has gained considerable attention. These qubits employ a layer of active error correction, significantly enhancing the encoded qubit lifetime and paving the way for more robust and reliable quantum computing systems.

3D superconducting resonators are not solely valuable for quantum computing, but they also have various significant applications. The outstanding features of 3D qubits make them powerful experimental testbeds for developments in quantum error correction[34] and bosonic quantum simulations[35,36]. Moreover, 3D qubits and their devices have many applications beyond QC, including the investigation of quantum phenomena under extreme conditions, quantum sensors, non-Markovian effects[37], quantum memory[30], fine materials loss measurements[38], cavity magnonics[39], and new fundamental physics, such as hidden sector photon[40] and dark matter[41].

The paramount importance of 3D qubits in high and ultra-high Q-factor cavities cannot be overstated, as evidenced by numerous review papers and tutorials[28,34,42–49]. However, the current literature lacks a comprehensive and systematic analysis that not only encapsulates the milestones achieved in this crucial domain but also delineates prospective avenues for future exploration. This glaring knowledge deficit serves as the catalyst for our present work, in which we strive to deliver an exhaustive review of the most recent breakthroughs in this domain. Our in-depth examination delves into the cutting-edge progress made in ultra-high Q-factor cavities, the integration of Josephson junction-based qubits, and the incorporation of BECs within 3D cavities. We meticulously scrutinize the origins of quantum state dephasing instigated by damping and noise mechanisms in both cavities and qubits, emphasizing the crucial challenges that must be surmounted to attain even greater coherence times. Furthermore, we present a critical assessment of the latest achievements in implementing single 3D qubits with superconducting materials and normal metals, as well as multi-qubit and multi-state quantum systems. Our work illuminates the prospects for the future of this research area, encompassing innovative materials for cavities and qubits, modes possessing nontrivial topological properties, advanced error correction strategies for BECs, and groundbreaking light-matter interaction effects.

## 2. Resonators and cavities

The core elements of circuit quantum electrodynamics (cQED) encompass a superconducting qubit intricately integrated within a MW resonator. In this section, we concentrate on the intricacies of resonators and cavities, while the subsequent section delves into the qubits and their interactions with resonators. Resonators can manifest in various configurations, such as 2D structures like superconducting coplanar-transmission-line resonators, lumped-element inductor-capacitor (LC) circuits, or 3D cavities. The so-called 2.5D resonators occupy an intermediate position within this hierarchy.

Regarding 2D transmission-line resonators, a qubit or a mumber of them are meticulously fabricated in close proximity to the resonator's central conductor. This conductor can feature strategically placed gaps, facilitating capacitive coupling with the resonator's input and output ports for enhanced performance[48,50]. This configuration results in normal modes with well-separated frequencies due to the open boundary conditions that arise when the current vanishes at these coupling points. To achieve a MW frequency in the range of 3-15 GHz (free-space wavelength of 10-2 cm), the resonator is typically designed with a length of several centimeters. These frequencies are high enough to prevent thermal photon population at the low operating temperature of ~10 mK, but still convenient for MW control electronics.

Lower frequency resonators can be characterized by the transmission line approach with the effective inductance ($L$) and capacitance ($C$). The resonant frequency and characteristic impedance of such a resonator are $\omega_r = 1/\sqrt{LC}$ and $Z_\mathrm{r} = \sqrt{L/C}$. The energy spectrum of the resonator is characterized by an equidistant ladder, with equally spaced transition frequencies $\hbar\omega_{nm} = \hbar\omega_r$ for any $n = m-1$. To observe this quantisation in practice one needs for the energy difference to exceed the thermal energy, $\omega_r \gg k_B T/\hbar$. Given that $1\mathrm{GHz} \times \hbar / k_B \approx 50\,\mathrm{mK}$, the condition can be readily met with MW-frequency circuits functioning at $\sim 10$ mK within a dilution refrigerator. Under these conditions, the Hamiltonian of this circuit takes the usual form for a quantum harmonic oscillator (QHO), $H_r = \hbar\omega_r a^\dagger a$. Here, the creation operator $a^\dagger$ can be expressed as $a^\dagger = \sqrt{1/2\hbar Z_\mathrm{r}}\left(\Phi - iZ_\mathrm{r}Q\right)$, where $\Phi$ is the flux threading the inductor and $Q$ is the charge on the capacitor with $[\Phi, Q] = i\hbar$. The operator $a^\dagger$ thus creates a quantized excitation of the oscillator's charge and flux degrees of freedom, or equivalently, of its electric and magnetic fields.

It is important to note that, even though the average field across the circuit is zero in the vacuum state, its root-mean-square value is non-zero: $E_\mathrm{rms} = d^{-1}\sqrt{\langle 0|V^2|0\rangle} = d^{-1}\sqrt{\hbar\omega_r/(2C)}$, where $V = Q/C$, and $d$ is the characteristic separation between conductors. For typical circuit parameters, $E_\mathrm{rms} d \approx 1$ μV. For the transmission-line resonator, where the distance between the center conductor and the ground plane is $d \approx 5$ μm, this corresponds to a zero-point electric field as high as $E_\mathrm{rms} \approx 0.2$ $\mathrm{Vm}^{-1}$. These large quantum fluctuations of the electric field arise from the small mode volume of the resonator and are among the reasons why the light-matter coupling in circuit QED can be much greater than in cavity QED[51].

Regarding 3D resonators, a diverse array of options is available for housing qubits in QED systems. These encompass rectangular, cylindrical, and spherical resonators, among other designs. Each resonator type exhibits distinct characteristics, such as mode volume and quality factor, which significantly impact the coupling strength between the qubit and the resonator. The optimal placement of a qubit within a resonator is contingent upon the distinct characteristics of the resonator in question. For instance, when employing a rectangular resonator, positioning the qubit at the electric field's antinode ensures maximum coupling strength. Conversely, in the case of cylindrical or spherical resonators, situating the qubit at the resonator's core, where the electric field is nullified makes the qubit remaining uncoupled from any specific resonator mode, which can offer substantial advantages for certain applications. The quantisation of modes of 3D cavities leads to the same Hamiltonian of a QHO, $H_r = \hbar\omega_r a^\dagger a$ .

In the case of 3D cavities, the root mean square of the vacuum electric field of the mode is defined by the mode volume, $E_{\mathrm{rms}} = \sqrt{\hbar\omega/(2\varepsilon_0 V_m)}$ . Here $V_m$ is the mode volume of the resonant cavity mode, $V_m = \int_V \varepsilon(\mathbf{r})\mathbf{E}^2(\mathbf{r})d\mathbf{r}/(\varepsilon(\mathbf{r}_q)\mathbf{E}^2(\mathbf{r}_q))$ . Here $\mathbf{r}_q$ refers to the qubit position inside the cavity, $V$ is the cavity volume. Significant coupling can be achieved through the utilization of larger mode volumes. This can be accomplished by utilizing larger transmon qubits, which compensate for any decrease in field amplitude[52].

Superconducting cavities offer exceptional coherence times and an infinite Hilbert space, both of which are vital for robust information storage - a fundamental requirement for Quantum Error Correction (QEC) and cavity-based quantum information processing. By employing high-Q 3D cavities, quantum data can be encoded in superpositions of the cavity's coherent photonic modes, such as the supremely coherent cat states. Owing to their intrinsically elevated quality factors, these qubits exhibit unparalleled coherence times, positioning them as a leading contender for realizing large-scale quantum computing.

## 3. Josephson and boson encoded qubits

Superconducting qubits can be primarily classified into two distinct categories: Josephson junction (JJ) qubits, which encode information in the supercurrents of circuits, and bosonic encoded qubits (BEQ), which store quantum information in the superposition of bosonic states of an oscillator. In this discussion, we will focus on the prevalent JJ qubits architecture that forms the backbone of contemporary quantum computers. JJ qubits employ a resonator, as described in the previous section, in conjunction with a nonlinear Josephson junction operating at temperatures below the critical threshold for superconductivity and in the $\omega_r \gg k_B T/\hbar$ limit that requires ~10 mK for $\omega_r$ in range 3-15 GHz[42,45,48,53]. Introducing JJ brings anharmonicity to the resonant circuit making the transitions between states $|0\rangle$ and $|1\rangle$ with corresponding frequency $\omega_{10} = \omega_1 - \omega_0$ uncoupled from other states.

The JJ typically consists of two bulk superconductors separated by a thin (~1-3 nm) insulating layer through which Cooper pairs can tunnel[54–56]. The supercurrent through the junction is described by the Josephson effect $I = I_c \sin(\phi)$, where the critical current $I_c$ is related to the Josephson energy $E_J$, $I_c = (2e/\hbar)E_J = (2\pi/\Phi_0)E_J$. Here, $\Phi_0 = 2\pi\hbar/(2e)$ is the magnetic flux quantum, $e$ is the charge of electron, and $\phi = \phi_L - \phi_R$ is the phase difference (Josephson phase) between the phases of the condensate wavefuction in the two superconductors across the junction. The time variation of the Josephson phase gives rise to the voltage drop $V$ across the superconductors, $d\phi(t)/dt = 2\pi V/\Phi_0$. For small currents $I \ll I_c$ the JJ can be described as an inductor with the intrinsic inductance ($L_J$) called Josephson inductance, defined by $L_J = \Phi_0/(2\pi I_c \cos\phi) = L_{J0}/\sqrt{1 - I^2/I_c^2}$, with $L_{J0}$ being an inductance at vanishing current[42,57].

The SQUID (Superconducting Quantum Interference Device) architecture, achieved by connecting two JJs in parallel, enables enhanced sensitivity to magnetic fields and tunable inductance, which is essential for effectively controlling quantum states of qubits within a superconducting circuit[45,46]. This configuration leverages the interference effects arising from the two junctions and contributes to improved coherence and stability in the quantum system, making SQUID-based designs a popular choice for various applications in quantum computing, sensing, and metrology. However, the introduction of this control knob comes at a cost, as the qubit's susceptibility to magnetic flux noise increases significantly.

The resonant frequency $\omega_{10}$ of a system can be conveniently adjusted to practical frequencies, 5-15 GHz, by using an additional capacitor $C_S$. The charging energy associated with the total system capacitance $C_\Sigma$ can be expressed as $E_C = q^2/2C_\Sigma$, where $C_\Sigma = C_J + C_S$. By increasing the value of $C_S$, the charging energy can be reduced sufficiently to achieve the low-charge noise operation regime ($E_J/E_C \gg 1$). This operation regime is essential for suppressing charge noise, which is currently the primary cause of rapid dephasing in JJ cQED devices. It also enables the realization of low-noise qubits, such as *transmons*[50,58,59], C-shunted flux qubits[60], and C-shunted fluxonium[32]. The inclusion of a shunt capacitor has resulted in longer coherence durations, falling within the 50 µs to 100 µs range. Nonetheless, this enhancement in the transmon's performance has come at a cost, as the anharmonicity has decreased to approximately ~200 MHz, representing only a small percentage of the qubit-level spacing[48]. Moreover, a very large shunting capactitance can make the transmon a physically large circuit.

Superconducting qubits are susceptible to dephasing, which can arise from charge noise or flux noise depending on the design of the qubit[61]. Charge noise results from fluctuations in the qubit's charge environment, which cause unpredictable changes in its charge state and impact its behavior. On the other hand, flux noise is caused by variations in the magnetic flux that passes through the qubit circuit, resulting in random fluctuations in the qubit's energy levels and introducing errors in quantum operations.

The energy and coherence decay times of superconducting qubits (JJ and bosonic) are characterized by $T_1$ and $T_\phi$. The total dephasing time $T_2$, during which quantum phase information is lost, is determined by both these lifetimes:

$$\frac{1}{T_2} = \frac{1}{2T_1} + \frac{1}{T_\phi}. \quad (1)$$

Here, the factor 2 before $T_1$ accounts for the fact that it describes the decay of field energy rather than field amplitude, as is the case for $T_2$ and $T_\phi$. One also introduces the corresponding decay rates, $\Gamma_1 = 1/T_1$, $\Gamma_2 = 1/T_2 = \Gamma_1/2 + \Gamma_\phi$. Notably, when there is no pure dephasing ($T_\phi^{-1} = 0$), the limitation for the damping lifetime is given by $T_2 = 2T_1$.

In the next section (see **Table 1**) we discuss various mechanisms of damping that contribute to the decay of $T_2$ and $T_\phi$ [62]. In the following section, we will delve into these mechanisms, while briefly presenting some key values associated with the most crucial qubits here. For the specific case of a 2D *transmon qubit* introduced by Schoelkopf et al. [50,58,59], ($E_J / E_C \gg 1$), with an anharmonicity of about 150-300 MHz and a weak sensitivity to charge noise, $T_1$ values can reach as high as 100-300 μs, while $T_2$ values can be around 10-200 μs for isolated planar qubits fabricated on a crystalline isolating substrate such as sapphire or Si. Placing a transmon inside a 3D cavity gives rise to the 3D transmon qubits. At the time of their invention in 2011, this approch enabled a significant growth in coherence times of the qubits up to $10^2$ μs and considered as the main approach to qubits with ultra high coherence times. There were several aspect that made us think this way. First, the high-Q resonator acts as a "quantum bus" that mediates the interaction between the qubit and the outside world. This interaction can be tailored to control the qubit and read out its state without causing significant dephasing [42,59,63–67]. Second, by coupling the qubit to a high-Q resonator, one can achieve strong coupling between them, which can be used to implement various quantum operations, such as two-qubit gates and to enable dispersive readout of the JJ-qubit states via a shift of the resonant frequency of the resonator. Nonetheless, an increasing number of subsequent investigations have demonstrated that the primary factors contributing to this advancement are enhanced fabrication techniques, refined materials, and suppression of electric fields in regions with high concentrations of noisy amorphous materials containing abundant TLS defects. Contemporary research indicates that utilizing BEQ qubits can facilitate even greater improvements in coherence times. The maximum coherence time achieved with these qubits is on the order of several thousand microseconds (~$10^3$ μs), primarily dictated by the 3D resonator's Q-factor.

Having discussed the essential aspects of qubits and resonators, we now focus on the qubit-resonator coupling. Under the assumptions of a Markovian and memoryless environment[68], weak system-environment interaction, and the initial absence of correlations between the system and environment, the Lindblad master equation formalism can be employed to describe the qubit-resonator interaction. The Jaynes-Cummings Hamiltonian for such a qubit-resonator system is given by[69]:

$$H = \hbar\omega_r a^\dagger a + \frac{1}{2}\hbar\omega_q \sigma_z + \hbar g(a^\dagger + a)(\sigma^- + \sigma^+), \tag{2}$$

where $\omega_r$ and $\omega_q$ represent the resonator and qubit frequencies, respectively, and $g$ denotes the resonator-qubit interaction strength (dispersionless around resonant frequency). Operators $a$ and $a^+$ ($\sigma^-$ and $\sigma^+$) are the annihilation and creation operators for the resonator mode (qubit). The characteristic coupling strength per photon, $g$, that enters Eq.(2) is given by $g = \mu E_{\mathrm{rms}} / \hbar$ [50,69]. Here, $\mu$ represents the expectation value of the dipole moment operator applied to the atomic state, and $E_{\mathrm{rms}}$ denotes the root mean square of the vacuum electric field of the cavity mode. In cQED, the interaction between components can be significantly incresed beyond what is achievable in atomic systems. As a result, we include both resonant and nonresonant terms in the Hamiltonian to thoroughly capture the strength of this interaction[51].

The master equation in Lindblad form for the reduced density matrix ($\rho$) can be utilized to describe the system:

$$\frac{\partial\rho}{\partial t} = -\frac{i}{\hbar}[H,\rho] + \gamma_r D[a]\rho + \Gamma_1 D[\sigma^-]\rho + \frac{\Gamma_\phi}{2} D[\sigma_z]\rho, \tag{3}$$

where $D[L]\rho = \frac{1}{2}(2L\rho L^\dagger - L^\dagger L\rho - \rho L^\dagger L)$ represents the Lindblad superoperator that describes the interaction of the resonator and qubit with the electromagnetic baths, leading to noise and damping[48,70–73]. The rate $\gamma_r$ signifies the decay of the resonator. This equation can be solved numerically using numerical quantum tools like QuTiP[72,74]. The model can be further generalized to account for the presence of higher energy levels[75,76].

In the case where both decay and dephasing of the qubit and resonator mode are neglected, and under the rotating wave approximation (RWA) which omits terms such as $a\sigma^-$, the Hamiltonian in Eq. (2) can be diagonalized to obtain the eigenvalues: $E_{\pm,n} = \hbar\omega_r(n+\frac{1}{2}) \pm \frac{\hbar}{2}\sqrt{4g^2(n+1)+\Delta^2}$, where $\Delta = (\omega_q - \omega_r)$ represents the resonator-qubit detuning and $n$ denotes the number of excitations (photons) in the resonator mode. When $\Delta \approx 0$ (small detuning), the eigenvalues exhibit a characteristic avoided crossing, referred to as Rabi splitting: $\delta E = E_{+,n} - E_{-,n} = 2\hbar g\sqrt{n+1}$. For $n = 0$, this phenomenon is also known as vacuum Rabi splitting. The interaction constant or photon number states in a superconducting circuit can be determined using this splitting for a given number of photons[77].

In the *dispersive coupling regime*[78], when the detuning between a qubit and a resonator is much grather than the coupling, $\Delta \gg g$, the Hamiltonian for the system is given by:

$$H \approx \hbar\left[\omega_r + \frac{g^2}{\Delta}\sigma_z\right]a^\dagger a + \frac{\hbar}{2}\left[\omega_q + \frac{g^2}{\Delta}\right]\sigma_z. \quad (4)$$

The Hamiltonian's resonator part acquires a resonance shift of $\pm g^2/\Delta$, which is dependent on the qubit state. It is important to note that this formula holds true only in the two-level approximation. In an experiment, the coupling strength can be found by measuring the value of frequency shift, $\chi = g^2\left[1/\Delta - 1/(\Delta-\alpha)\right]$, depending on the qubit state, $\alpha$ denotes the difference in the transition frequencies caused by the anharmonicity of the JJ. The state-dependent shift enables qubit state measurements, and the dispersive regime is commonly employed in quantum computing for implementing high-fidelity two-qubit gates and qubit state measurements.

## 4. Dephasing and noise in resonators and qubits

Despite being in the superconducting regime, cavities and qubits experience dissipative material losses, restricting the intrinsic Q-factor. Before delving into the various loss and dephasing mechanisms, it is instructive to determine the maximum possible coherence values based on the quality of materials accessible through cutting-edge fabrication. We can then contrast these with the reported coherence times for JJ and BEC qubits. For example, the Q-factor of sapphire resonators in the GHz frequency range can be very high, often reaching values on the order of $10^7$ to $10^9$ or even higher, depending on the experimental conditions and quality of the sapphire corresponding to the coherence times value exceeding 30,000 μs[79–82]. In superconducting 3D microvave resonators these Q-factors and times are very eay to achive due to the concentration of E-fields in the free space rather than in material as in the dielectric resonators[30,31,67,83–85]. For 3D resonators recording values of Q-factor are around $10^{10}$ at 5 GHz, that correspond to the decay times of a few seconds (~$10^6$ μs)[86].

What limits the values of intrinsic Q-factor ($Q_i$) of resonators? Intrinsic Q-factors of resonators at cryogenic temperatures are limited by various factors that contribute to the overall energy loss and dephasing. These factors can be classified into internal and external factors. At cryogenic temperatures, internal factors become dominant due to the reduction in thermal noise. As shown in a series of works[84], Q-factor limitations depend on dielectric losses from the oxide layer lining the interior of a cavity and damping from finite residual surface resistance. Although MW engineering allows mitigating the loss to a certain extent by controlling the spatial distribution of the particular resonant mode, they cannot be removed completely. At cryogenic temperatures, surface losses become more pronounced as they are less affected by temperature reduction compared to other loss mechanisms. The part of the losses associated with the material rather than the electromagnetic mode is associated with two main aspects: *excess quasiparticles* (QPs)[87,88] and *parasitic two-level systems* (TLS)[38,89,90]. TLS are microscopic defects that can be present in dielectrics or amorphous materials. They can absorb and re-emit energy from the resonator, causing energy dissipation. At cryogenic temperatures, TLS can be a dominant source of loss, limiting the Q-factor. Finally, while radiation losses and nonlinear effects may be less critical for cryogenic resonators and qubits, they still hold the potential to impact intrinsic Q-factors. Radiation losses, although generally minimal at cryogenic temperatures, can still contribute to the overall energy loss in the system. This becomes particularly significant when the resonator lacks adequate shielding or if the design permits radiation leakage. Furthermore, nonlinear effects can impose limitations on a resonator's Q-factor. At cryogenic temperatures, these effects may stem from the nonlinear behavior of materials or from interactions with other devices within the system. As such, it is vital to consider these factors when optimizing the performance of cryogenic resonators and qubits.

Discerning microscopic loss mechanisms presents a significant challenge. In **Table 1**, we provide a comprehensive overview of various loss, dephasing, and noise mechanisms, accompanied by proven strategies for mitigating their impact. In this section, we delve deeper into the primary dephasing and noise mechanisms to offer a more thorough understanding.

**Table 1.** Various damping mechanisms in superconducting cavities and qubits, along with their mitigation methods.

| Damping and noise source | Mitigation | Refs. |
|---|---|---|
| *Superconductor Residual AC Surface Impedance*: The residual AC surface impedance in superconductors can contribute to energy dissipation in superconducting cavities and qubits | Vacuum heat treatment; removing grains and grain boundaries | 91,92 |
| *Finite dielectric loss of wall oxide layers*: Oxide layers on the walls of superconducting devices can lead to dielectric losses | Modes with vanishing electric field at the surface (e.g., TE011 mode in a cylindrical cavity); Reduction of surface participation ratio and filling factor; The use of materials with lower losses (e.g., Ta) | 84,93,94 |
| *Conductivity and defects in qubit substrate*: Imperfections in the substrate material can affect the qubit's performance | Using materials with ultralow conductivity (sapphire, c-Si); suspending superconducting qubits | 95 |
| *Edge defects*: The edges of superconducting devices can harbor defects that impact the quality factor of resonators and qubits | Smoothing the topology of the resonator; Avoiding edges, e.g., concentric 2D transmon, the cylindrical 3D transmon | 96 |
| *Photon number fluctuations*: Fluctuations in the photon number within the resonator can contribute to the dephasing of qubits | Lower temperatures; materials with high thermal conductivity for thermalization; suppression of coupling with input and output ports with attenuators; dynamical decoupling | 97 |
| *Two-level-system (TLS) defects*: TLS defects, typically caused by impurities or structural defects in materials, can lead to energy dissipation and decoherence in superconducting devices | Suppression of filling factor in the areas of high TLS; Proper surface treatments like removing native oxides, etching and passivation; Avoiding amorphous materials; | 38,89,98,99 |
| *Nonequilibrium quasiparticles*: The presence of nonequilibrium quasiparticles can result in relaxation and dephasing of qubits | Quasiparticles trapping by vortices, pumping pulses and normal metal traps | 100–103 |
| *Purcell decay*: A reduction of $T_1$ due to the accelerated spontaneous emission when a qubit is coupled to a resonator | Qubit frequency tuning; reducing the coupling qubit-cavity; use a Purcell filter | 104–106 |
| *Radiative losses*: These losses result from the emission of electromagnetic radiation by the superconducting device | Selection rules; the use of dark (forbidden) states | 107 |
| *Photon shot noise and random AC-Stark shift:* Fluctuations in cavity photon number can cause shifts in the qubit frequency, leading to dephasing | Operating at lower cavity photon numbers; strong dispersive coupling regime; use of a dynamical decoupling protocol | 61,108 |
| *Charge noise*: Fluctuations in the charge environment can affect the qubit frequency and coherence | Operate at the charge-insensitive point; improved fabrication processes; advanced materials and designs; larger effective junction capacitance; careful material selection; proper device design; the use of charge-insensitive qubits | 58,61,109 |
| *Flux noise*: Magnetic flux fluctuations within a superconducting loop can impact qubit performance, especially in flux-tunable qubits | Operate at the flux-insensitive point; Reduction of the loop size in a flux-qubit circuit; magnetic shielding | 61,110 |
| *Interaction with higher energy levels*: Higher energy levels in a qubit system can lead to unwanted transitions and decoherence | Higher anharmonicity; excitation and readout pulse engineering; selection rules; implementing dynamical decoupling techniques; designing qubits with reduced susceptibility to higher energy level interactions | 111,112 |
| *Readout Noise*: The process of measuring a qubit's state can introduce noise, which can affect the accuracy and reliability of quantum computations | Using quantum-limited amplifiers, cryogenic amplifiers; employing error-correction techniques to improve the readout fidelity; new readout approaches, e.g., direct measurement at the millikelvin stage | 113–116 |

The primary challenge in achieving high Q-factors at millikelvin temperatures and single-photon energies lies in understanding and mitigating the impact of parasitic TLSs that lead to decreased intrinsic quality factors ($Q_i$)[38,89,117–120]. TLSs are inherent in any material due to finite purity in bulk and at the boundaries. Various experiments have investigated the role of resonant TLS in introducing noise to a resonator[119,120]. It has been demonstrated that TLS can interact with each other[121,122], causing them to move in and out of resonance with a superconducting device. This fluctuation serves as a source of noise, as evidenced by the characteristic temperature dependence of 1/f noise[99,119,123–125]. This effect was directly observed in a 3D-transmon circuit[122], which resulted in a time-varying qubit relaxation rate.

One common approach to minimize the impact of TLS is to avoid using amorphous materials. However, despite substantial research efforts and improvements in Q-factors, it remains impossible to completely eliminate parasitic TLS. The scientific community has instead developed strategies to circumvent this issue, such as employing 3D cavities where the E-field is localazed reliminary in the free space minimasing the participation ratio of the surface. The *participation ratio* of a resonator refers to a metric that quantifies the extent to which various regions or materials in a resonator structure contribute to the total energy stored in the resonator. In the context of superconducting resonators and qubits, the participation ratio is often used to evaluate the influence of different materials or interfaces on the overall energy dissipation and loss mechanisms, such as those caused by TLSs. A higher participation ratio for a specific material or region within the resonator implies a greater contribution to the overall energy stored in the resonator. Conversely, a lower participation ratio indicates a smaller contribution. By analyzing participation ratios, one can identify and understand the dominant sources of energy loss in a resonator or qubit system, and then optimize the design or choice of materials to minimize these losses and improve the device's performance. In the case of resonators with multiple materials or interfaces, it is important to minimize the participation ratio of materials that are prone to energy dissipation, such as amorphous materials, which can host a higher density of TLS[99]. By doing so, the resonator's overall energy loss can be reduced, leading to a higher quality factor and better performance. For example, for larger structures, the participation ratio of amorphous materials at interfaces is reduced, further mitigating the impact of TLS on resonator performance.

Nonequilibrium quasiparticles (QPs) are essentially unpaired electrons that arise from the breaking of Cooper pairs due to temperature fluctuations and external electromagnetic radiation. These QPs can adversely impact the performance of superconducting qubits by causing relaxation and dephasing, with the rate being approximately proportional to the QP density[126]. Within a superconducting device, the qubit's degree of freedom can exchange energy with QPs, resulting in an intrinsic relaxation mechanism. This mechanism is suppressed in thermal equilibrium at temperatures significantly lower than the critical temperature, owing to the exponential depletion of the quasiparticle population. However, in superconducting qubits and resonators, the presence of nonequilibrium quasiparticles has been observed, leading to relaxation even at millikelvin temperatures[126–128]. There are several strategies to address the issue of nonequilibrium QPs, including quasiparticle trapping using vortices[100], pumping pulses[101], and normal metal traps[102]. High-quality-factor 3D cavities serve as a powerful tool for investigating the dynamics of quasiparticles in superconducting qubits[100], offering valuable insights for enhancing the performance and coherence of quantum devices.

One effective approach to mitigating losses in resonators and qubits involves exploring alternative materials beyond traditional aluminum (Al) and niobium (Nb). For instance, tantalum (Ta) has been demonstrated to achieve significant improvements in 2D transmon qubits when used as the superconductor in capacitors and MW resonators[94,129]. This enhancement is attributed to the insulating oxide of Ta, which reduces MW loss in the device[130]. Investigating substrate materials with lower imperfection densities, resulting from unique surface chemistries, presents another promising avenue for enhancing the coherence of superconducting quantum devices. In a recent study[131], two ternary metal oxide materials, spinel ($MgAl_2O_4$) and lanthanum aluminate ($LaAlO_3$), were investigated. Devices fabricated on $LaAlO_3$ exhibited quality factors three times higher than those of previous devices, due to a reduction in interfacial disorder. Interestingly, although $MgAl_2O_4$ demonstrated significant surface disorder, it consistently outperformed $LaAlO_3$, positioning itself as a promising novel material for superconducting quantum devices. This suggests that exploring diverse materials with distinct properties can lead to substantial improvements in the performance and coherence of superconducting qubits and resonators.

In both superconducting cavities and qubits, their performance and coherence are impacted by numerous other mechanisms that cannot be entirely resolved by enhancing material quality or employing state-of-the-art fabrication techniques, as illustrated in **Table 1**. Radiative losses result from the emission of electromagnetic radiation by the superconducting device, while photon shot noise and random AC-Stark shift arise from fluctuations in cavity photon number, causing shifts in the qubit frequency and leading to dephasing. Charge noise stems from fluctuations in the charge environment, which can impact qubit frequency and coherence. Flux noise, on the other hand, originates from magnetic flux fluctuations within a superconducting loop, affecting qubit performance, particularly in flux-tunable qubits. The interaction with higher energy levels in a qubit system can lead to unwanted transitions and decoherence. Readout noise is introduced during the process of measuring a qubit's state, which can compromise the accuracy and reliability of quantum computations.

Finally, we delve deeper into the Purcell decay, which is particularly relevant to the subject of this work, namely, qubit-cavity interactions. The Purcell decay refers to the decrease in qubit lifetime resulting from accelerated spontaneous emission when a qubit is coupled to a resonator. The longitudinal relaxation time $T_1$ can be altered due to changes in the spontaneous emission rate of a qubit in the presence of a cavity, a phenomenon known as the Purcell effect. Named after Edward Purcell, who first observed it in 1946[132], the Purcell effect bears significant implications for quantum information processing and quantum communication. It can limit qubit coherence time and reduce the fidelity of quantum operations. To increase $T_1$ by mitigating the Purcell effect, one can employ engineered cavities that suppress certain electromagnetic modes particularly detrimental to qubits or use high-quality qubits that are less susceptible to the Purcell effect. Conversely, rapid energy emission can be crucial for certain quantum operations. By adjusting the coupling strength between the qubit and the cavity, the Purcell effect can be optimized, resulting in higher fidelity quantum operations.

In the weak-coupling regime, the hybridization of qubit and resonator eigenstates is weak, so the emission frequency $\omega_q$ remains unaltered by the resonator. In other words, the Lamb-shift is significantly small compared to the original resonant frequency, and the light-matter interaction only modifies the decay rate. The qubit transition dipole moment is denoted by $\boldsymbol{\mu}$. The ratio of the decay time in free space $T_1^0$ to the decay time of the same qubit in the resonator $T_1$ can be expressed as[133,134]: $F \equiv \frac{T_1^0}{T_1} = 1 + \frac{6\pi\varepsilon_0}{|\boldsymbol{\mu}|^2} \frac{1}{q^3} \mathrm{Im}[\boldsymbol{\mu}^* \cdot \mathbf{E}_\mathrm{s}(\mathbf{r}_\mathrm{q})]$, where $q = \omega / c$ is the wavenumber in free space, $\mathbf{E}_\mathrm{s}(\mathbf{r}_\mathrm{q})$ is the scattering part of the electric field evaluated at the qubit position $\mathbf{r}_\mathrm{q}$, and the qubit has a dipole moment $\boldsymbol{\mu}$ oscillating at the frequency $\omega_q$. The quantity $F$ is called the Purcell factor, and this formula holds for any resonator.

In the single-mode approximation and in the absence of other damping mechanisms, the coupled-resonator and qubit system exhibit a maximal Purcell-limited lifetime defined by[30,47,60,66]:

$$T_{1,\mathrm{Purcell}} = \frac{\Delta^2}{\kappa g^2}, \tag{5}$$

where $\Delta$ is the cavity-qubit frequency detuning, $\Delta = \omega_c - \omega_q$, $\kappa$ is the total loss rate (spectral width in the Lorentzian spectral shape), and $g$ represents the qubit-cavity coupling strength. For a low-loss system, $T_{1,\mathrm{Purcell}}$ sets the upper limit for the energy ($T_1$) and coherence ($T_2$) lifetimes: in the absence of pure dephasing, the coherence lifetime is $T_2 \approx 2T_1$. Therefore, to enhance coherence and relaxation times, one must suppress the Purcell effect by employing qubit frequency tuning, reducing the qubit-cavity coupling, and utilizing a Purcell filter[104–106].

In the strong-dispersive regime, a qubit's sensitivity to stray cavity photons is significantly increased, leading to dephasing as a result of their random ac-Stark shift[65,135–137]. This dephasing can be mitigated in cavities possessing a higher Q-factor. As demonstrated in Ref.[137], enhancing the cavity mode decay time results in a substantial improvement of the 3D qubit coherence time $T_2$ in the dispersive coupling regime, with an increase of up to three times (from 10 μs to 25 μs in the referenced study). The qubit relaxation time remains relatively unaffected, owing to the strong detuning and the consequently suppressed Purcell effect.

The frequency noise spectrum of a superconducting resonator and JJ qubits at a few and single photon level is described by various noise processes, including fluctuations in the resonator's physical parameters, interactions with the environment, and the presence of TLS. These processes contribute to the overall noise in the system, which can manifest as 1/f noise, white noise, and other forms of noise spectra, depending on the dominant loss mechanisms and the operating temperature of the device. The frequency noise of a superconducting resonator at a few[99,119,138], and single[139] photon energies is well described by 1/f dependence, $S_{1/f}(f) = h_{-1}/f^{\alpha} + h_0$, where $h_0$ is the white frequency noise level, $h_{-1}$ is a flicker frequency noise level and $\alpha$ is an exponent describing the strength of low-frequency noise components. This ubiquitous type of noise has been assigned mainly to TLSs in amorphous materials (e.g., oxides at the surfaces and nm-thick dielectric of the JJ).

## 5. Advances in 2D qubits

Before delving into the diverse 3D cavity and 3D qubit configurations, it is beneficial to briefly recapitulate the progress made in 2D qubits. The recent advances in 2D qubits that appeared over the last 5 years include improvements in qubit coherence times, the development of novel error-correction techniques and the emergence of new materials and fabrication methods. These advancements have paved the way for more scalable and reliable quantum devices, and have contributed significantly to the progress of the field.

Researchers have reported significant improvements in the coherence times of 2D qubits, which directly impacts the performance of quantum computers. IBM has made remarkable progress in the field of 2D transmon qubits, achieving significant breakthroughs recently[140–142]. As a result of IBM's state-of-the-art fabrication techniques, their cloud-based quantum computers have qubits with $T_1$ and $T_2$ times ranging from 100 µs to 300 µs, which are exceptionally long values. Such transmons are not flux tunable. Nevertheless, IBM has unveiled their newest processor, the "IBM Osprey," which consists of 433 qubits, making it the most extensive quantum processor produced by the company to date[143].

Significant progress has also been made in the development of qubit types beyond transmon. Recently, the design and fabrication of flux qubits have been revisited[61]. Reported relaxation and coherence times of 40 µs and 85 µs respectively, indicate enhanced performance. The planar device exhibits broad-frequency tunability, strong anharmonicity, and high reproducibility. Moreover, residual thermal photons primarily contribute to qubit dephasing at the flux-insensitive point in the readout resonator. Ref. [144] examined the airbridge topology in superconducting MW circuits to mitigate coupling to parasitic modes in coplanar waveguides. Although the predicted intrinsic quality factor $Q_i \approx 6.3\times10^5$ is high, the experimental values were found to be lower, suggesting a need for further exploration of this approach. The adoption of alternative capacitor geometries, such as Xmon[145], has gradually and collectively improved coherence times to the 50-100 µs range[145,146] and beyond[147,148]. Notably, record-breaking $T_1$ and $T_2$ times of ~0.5 - 1.5 ms have been achieved in the 2D topology using fluxonium[32,33].

The search for new materials and fabrication methods has also led to the discovery of novel 2D qubit platforms. For example, exploring new materials beyond Al and Nb, as well as designs that have achieved record quantum coherence times. In Ref. [149], it was demonstrated that utilizing TiN in place of Al for fabricating 2D interdigitated shunting capacitor-based transmon qubits leads to a significant enhancement in qubit coherence times, with reported values of $T_1 \approx T_2 \approx 55-60$ μs (qubit quality factor $Q \approx 2\times10^6$). These values are comparable to the lifetimes of existing 3D transmons and surpass the typical values for Al 2D transmons by a factor of 2-3. A description of the employed technological process is instructive. The 2D transmons were fabricated from TiN grown on high-resistivity Si substrates, which were prepared using an HF etch to eliminate the native oxide. The hydrogen-terminated surface remained stable long enough for transfer to the deposition system. The substrate was heated to 500°C, and then a Ti target was reactively DC sputtered using a mixture of Ar and $N_2$ gases. The resulting Ar:$N_2$ plasma formed a thin (2 nm) amorphous SiN layer on the substrate. Subsequently, a 30 nm thick film of TiN was deposited using reactive sputtering from a pure Ti target in the Ar:$N_2$ atmosphere at the same temperature. The qubits were fabricated by depositing Nb regions in non-device areas for alignment purposes, patterning the wafer with contact lithography in i-line resist, and conducting a timed $Cl_2$/$BCl_3$ reactive ion-etching at low pressure and high bias. The etching resulted in a 125 nm recess in the Si substrate. Immediately after etching, the remaining resist was stripped through manual agitation in an acetone bath, and chlorine (Cl) residue was removed with a room temperature water bath. The JJs were fabricated using standard shadow evaporation techniques.

## 6. Superconducting cavities for JJ and bosonic qubits

Despite their ease of fabrication, scalability, and compactness, two-dimensional (2D) resonators are constrained by significant surface dielectric losses stemming from the concentration of electromagnetic energy near the substrate surface. These resonators are plagued by various loss mechanisms, such as dielectric losses at the superconducting material-substrate interface, excess quasi-particles, oxide and contaminant-induced losses, and surface spins and charge noise that contribute to dissipation and dephasing[150]. In 2011, researchers showed that some dephasing mechanisms could be mitigated in three-dimensional (3D) transmon qubits due to their larger volume and lower field concentration in areas with high concentrations of TLS centers[52]. 3D resonators achieve longer coherence times, higher stored energy, and significantly reduced surface losses, as the electric field is predominantly stored in a vacuum. With internal Q-factors often surpassing $10^8$ [84], 3D cavities outperformed their 2D counterparts at the time by two orders of magnitude[151] and have become one of the main platforms for cQED[152,153].

Not only do 3D cavities offer high-Q modes, they also present various other advantages for quantum systems engineering. These resonators are expertly crafted to provide ultimate protection for qubits against external noise sources, such as electromagnetic radiation and thermal fluctuations[52]. Furthermore, 3D resonators offer improved isolation between qubits, significantly reducing the likelihood of crosstalk and undesired qubit interactions. This heightened isolation effectively safeguards the integrity of quantum states, consequently prolonging the coherence time of qubits. The mode structure of a 3D resonator can be meticulously designed to achieve the optimal balance between coupling strength and environmental isolation, thereby enhancing their performance for quantum applications. Lastly, 3D high-Q resonators enable the BECs and various approaches of quantum error correction.

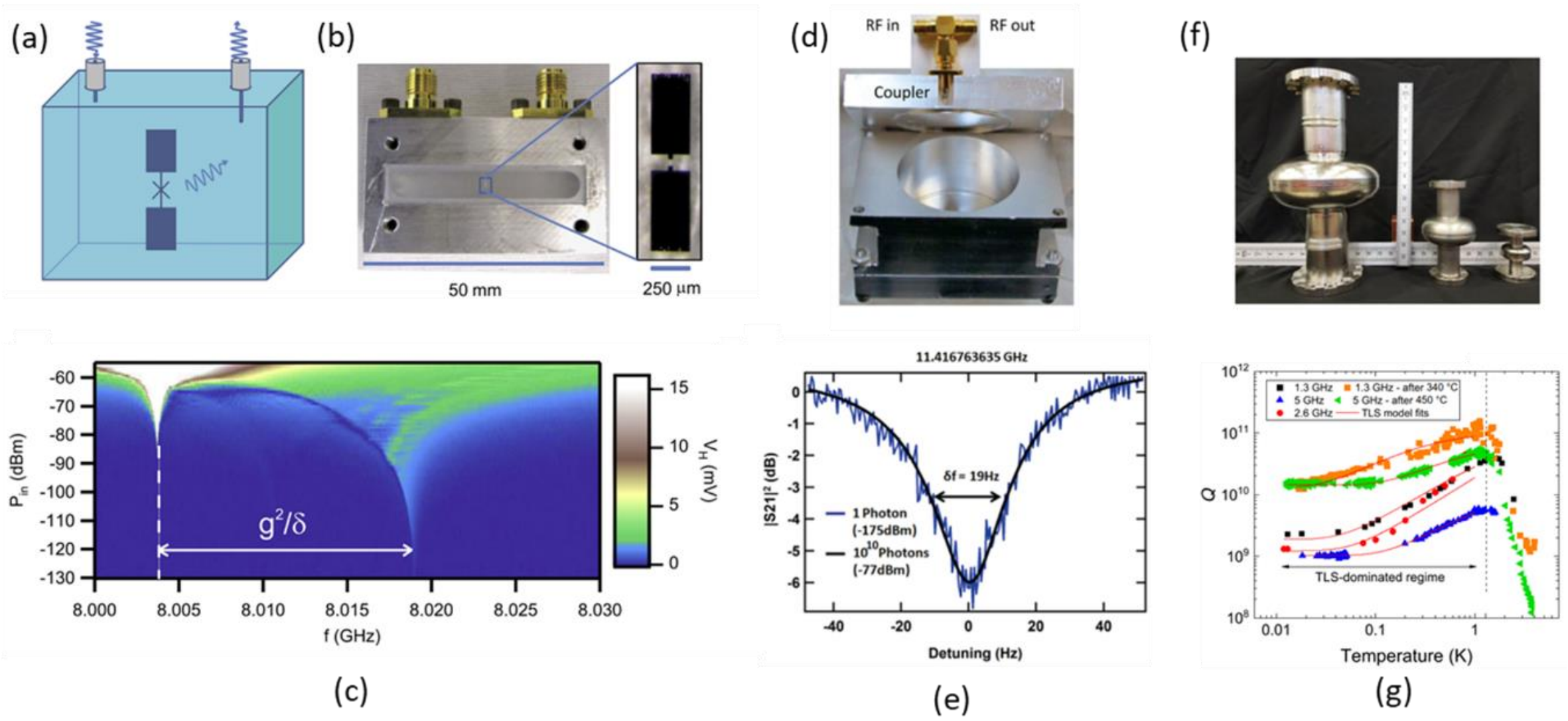

**Figure 2**. **Single 3D qubits made of superconductive materials.** (a) Transmon qubit inside a 3D cavity, coupled via a broadband dipole antenna for photon reception and emission. (b) Half of 3D Al cavity. An Al transmon qubit with a dipole antenna is mounted on a c-plane sapphire substrate at the cavity center. Insert: Optical microscope image of a single-junction transmon qubit. Dipole antenna length: 1 mm. (c) 3D cavity coupled to transmon measured by power and frequency. Cavity frequency shifts by $g^2/\delta$, where $\delta = \Delta$ is the frequency detuning[52]. (d) Schematic of a cylindrical resonator measured in shunt configuration. (e) Single-photon (blue) and high power (black) transmission data of the configuration are shown left. The best fit at single-photon power is indistinguishable from the high power trace, showing the remarkable power-independence of the TE011 mode[84]. (f) Single-cell Nb cavities of the TESLA geometry. (g) Intrinsic cavity quality factor Q of the 1.3-, 2.6-, and 5-GHz TESLA cavities as a function of temperature[86]. Q reaches its maximum around 1K and then decrese upon themperature redusing due to TLS.

Superconducting 3D resonators with very high Q-factors enable storage times approaching seconds[154,155], even at the single-photon level[156]. Such cavities are commonly used to trap and store resonant

MW radiation and reduce losses, allowing devices with very high Q-factor, narrow bandwidth, and long storage times[157]. Such cavities are essential for many physics applications, including particle accelerators[158], ultra-sensitive motion/displacement sensing[159], precise frequency stabilization[160], and testing fundamental physics. In particular, tests of the speed of light and the constancy of fundamental constants[160–162], as well as the search for hidden sector particles, which are dark matter candidates[163,164], rely on such cavities.

In the pioneering work[52], a remarkable enhancement in the single-qubit relaxation lifetime $T_1$ and coherent lifetime $T_2$ was achieved, with values of 60 μs ($Q_1 \approx 2\times10^6$) and 10-20 μs ($Q_2 \approx 7\times10^5$), respectively. At the time, this improvement was more than an order of magnitude compared to the 2D counterparts. It was explained by the fact that the larger mode volume of the cavity led to a reduced E-field strength at the metallic components, resulting in lower dissipations due to the residual AC resistance of the superconductive state. The $T_2$ values were found to be less than the $2T_1$ limit, suggesting the presence of substantial pure dephasing in the system. The 3D cavity was fabricated from superconducting Al alloy (6061 T6) as shown in **Fig. 2a**. A strong interaction between the transmon and the lowest mode (TE101) of the suppressed field was achieved using a broadband dipole antenna with an increased size, which was connected to the Josephson junction (JJ) for receiving and emitting photons as depicted in **Fig. 2a**. The realized 3D transmons exhibited vacuum Rabi frequencies ($g/2\pi$) greater than 100 MHz. The larger antenna size contributed to a higher shunting capacitance, thus reducing charge noise. Moreover, the transmon's position ensured minimal coupling to the second (TE102) cavity mode. In the dispersive limit ($\Delta\omega \gg g$), where $\Delta\omega$ represents the frequency detuning between the transmon transition frequency and the cavity resonance, the 3D transmon exhibited a strong dependence of the resonance shift on the input power ($P_{in}$), enabling dispersive readout of the qubit state as shown in **Fig. 2c**. The reported result of $T_2 < 2T_1$ and the residual dephasing were ascribed to the non-$1/t$ type of noise associated with the generation of QPs. This was further corroborated by the observation of a sudden increase in dephasing when the temperature exceeded 130 mK.

In-depth research has revealed multiple crucial factors that can substantially boost the Q-factor of cavities. The first factor involves surface treatment, which has proven to yield remarkable enhancements in resonators. For instance, in Ref.[151] it was showed that meticulous surface preparation prior to Al deposition on sapphire can achieve Q-factors surpassing $10^7$ at high powers ($10^6$ in the single-photon regime), by generating exceptionally smooth and pristine interfaces between the Al metallization and the underlying single crystal sapphire substrate. Secondly, recent discoveries highlight the importance of interfaces between distinct components of 3D cavities in defining their quality factors[81]. Thirdly, a comprehensive investigation into surface participation and dielectric loss in 3D JJ qubits has demonstrated that the qubit relaxation rate $T_1$ is directly proportional to the surface participation ratio, given a clean electromagnetic environment and reduced nonequilibrium QP density[93]. This emphasizes the crucial role played by the finite conductivity of cavity walls when other loss sources are absent. Lastly, considering that a 3D qubit can only be approximately represented as a two-level system, it is imperative to take into account the coherence and decay of the qubit's higher energy levels[111].

In 2013 Matthew Reagor et al. [84] achieved a significant milestone by attaining a 10 ms single-photon lifetime in superconducting Al cavities. They investigated both rectangular and cylindrical cavity geometries, finding that cylindrical cavities offered superior performance. Specifically, chemically-etched Al (5N5) cylindrical resonators displayed a 10.4 ms photon lifetime, vastly outperforming their rectangular counterparts at 1.2 ms, as shown in **Fig. 2d**. The cylindrical cavity's advantage was credited to its TE011 mode, which eliminates electric energy storage at cavity walls. The team successfully demonstrated reproducible internal quality factors exceeding $0.5\times10^9$ in both single-photon and multi-photon excitation, **Fig. 2e**. The cavities were fabricated from bulk Al with purity levels ranging from 95% (alloy 6061-T6) to 99.9995% (5N5). Surface treatments involved immersing the cavities in a phosphoric-nitric acid mixture. Consistent with findings on Nb resonators[154,157], removing 100 µm of material yielded the longest-lived resonators in pure, bulk Al. Chemical etching and annealing processes for 3D Al cavities resulted in high internal quality factors on the order of $10^8$ at single-photon levels. While coupling with a transmon might reduce $Q_i$ this ultrahigh Q resonator holds potential for quantum memory applications.

These results and the loss analysis in 3D cavities (**Table 1**) suggest that higher Q-factors are achievable in cylindrical cavities without edges. These cavities support modes localized within the cavity volume, resulting in reduced surface material participation factors. Ref.[67] further investigates this concept, exploring an optimized 3D coaxial cavity design (*RadiaBeam*). This optimized cavity exhibits an impressive intrinsic $Q_i \approx 10^7$, a 25% improvement over non-optimized designs.

TESLA-shaped elliptical cavities, made from high-purity bulk Nb and fabricated through deep drawing and electron beam welding, are of particular interest for particle accelerators due to their exceptionally high intrinsic Q-factors[129]. Recent work by Romanenko et al. [86] explored single-cell TESLA Nb cavities at low photon excitation regimes and temperatures of ~10 mK (**Fig. 2f**). Intrinsic Q-factors reached an astounding $4\times10^{11}$ at 1.4 K after proper surface treatments, with empty cavity (no coupled qubit) photon lifetimes up to 2 s (**Fig. 2g**). The TM010 mode of this axially symmetrical cavity, with minimal surface participation, was employed. Upon cooling to mK temperatures, the cavity exhibited a tenfold decrease in Q-factor, attributed to TLS contribution. TLS saturation occurs at ~1 K temperatures and multi-photon excitation, becoming evident in low-photon excitations and temperatures.

3D Nb-based superconducting cavities for particle accelerators are designed for high RF power, achieving performance close to theoretical limits. This remarkable accomplishment stems from the meticulous design and processing of cavity interior surfaces. Typical surface finishing techniques, such as buffer chemical polishing and electropolishing, minimize surface contamination and roughness. High-temperature heat treatment steps provide mechanical stability and hydrogen degassing, preventing quality factor degradation due to the formation of normal conducting Nb hydride precipitates on cavity surfaces. State-of-the-art cavity processing techniques include impurity doping (Ti, $N_2$, $O_2$) to reduce residual loss by modifying RF surfaces[165–168]. While these performance improvements have been primarily observed in high RF fields and operating temperatures (~2.0 K), which are above the interests of quantum communities, they still offer valuable insights for further development.

Superconducting cavities in modern 3D qubit systems are typically formed[86,169] or precision-machined[47] from ultra-high-purity aluminum or niobium, which demands significant cost and time for optimal surface preparation. However, alternative superconducting materials exist, such as the 6061 aluminum alloy with a maximum impurity composition of 0.8% silicon, 0.7% iron, 0.15% copper, and 1.2% magnesium. This alloy has demonstrated superconductivity with Q-factors reaching the order of $10^{684}$ and has been employed in large-scale acoustic resonant-mass gravitational wave detectors[170].

The emergence and growth of 3D printing (also known as additive manufacturing or AM) in recent years has revolutionized the production of metallic components, making it faster and more cost-effective. AM technologies fabricate parts layer by layer using computer-aided design (CAD) data, offering unparalleled geometric and material design freedom. Sames et al. [171] provide an in-depth review of metal AM science. Powder Bed Fusion AM, which utilizes a focused laser or electron beam to fully melt metallic powder on a flat bed, has been employed to create radio frequency (RF) resonant cavities from various materials. Electron beam-based Powder Bed Fusion (EB-PBF) is particularly suitable for unalloyed copper[172] and niobium RF components[173,174], where copper's high reflectivity and thermal conductivity or niobium's high melting temperature present challenges for many low average power laser-based Powder Bed Fusion (L-PBF) systems. Operating in a vacuum environment as low as $10^{-7}$ Torr, EB-PBF better preserves the original powder feedstock purity. Although L-PBF systems offer superior resolution and surface roughness compared to EB-PBF, they are primarily limited to metallic alloy systems based on titanium, aluminum, nickel, cobalt, and iron.

In Ref.[174] quarter-wave resonators (QWR) were 3D printed using EB-PBF and L-PBF from reactor-grade niobium (Nb, RRR30) and Ti-6V-4Al (Ti64), respectively. **Table 2** displays the results of both room temperature and cryogenic measurements, comparing 3D printed and precision-machined QWRs with identical designs[67].

**Table 2:** Room temperature and cryogenic RF measurements of 3D printed QWRs (adapted from [174,175]).

| # | Material | Fabrication method | Warm | | | Cold (4K) | |
|---|---|---|---|---|---|---|---|
| | | | $Q_0$ Expected | $Q_0$ Measured | $f_0$ (MHz) | $Q_0$ | $f_0$ (MHz) |
| | Copper (C10100) | Machined | 3492 | 3312 | 6027.7 | - | - |
| | Nb (RRR 300) | Machined, Etched (-89µm) | 1190 | 1069 | 6045.5 | $9.2x10^6$ | 6192.9 |
| 1 | Nb (RRR 30) | EB-PBF, abrasive polish, and etched (-10µm) | 1190 | 425 | 5991.8 | $1.6x10^6$ | 6008.7 |
| 2 | Nb (RRR 30) | EB-PBF, abrasive polish, and etched (-10µm) | 1190 | 487 | 6113.2 | $1.2x10^6$ | 6130.4 |
| 3 | Ti64 | L-PBF, HIP, tumbled, etched (-10µm) | 622 | 345 | 5962.6 | $0.4x10^6$ | 5980.5 |
| 4 | Ti64 | L-PBF, HIP, tumbled, etched (-10µm) | 622 | 296 | 5894.4 | $2.2x10^6$ | 5917.1 |

Examining **Table 2**, we observe a strong correlation between the warm temperature RF measurements of precision-machined copper and Nb QWRs and their anticipated Q values. Additionally, the center frequencies were measured to be within 1% of the target frequency (6 GHz). In contrast, the AMed QWRs demonstrated a broader frequency range, attributable to the geometric accuracy and surface finish variations stemming from the 3D printing process. In terms of Q, the Nb AMed cavities reached 35% and 40% of the ideal value. Similarly, the abrasively finished Ti64 AM cavity achieved 55% of the ideal Q, while the etched cavity reached 47%. Despite the absolute Q of Nb AM cavities surpassing that of the Ti64 AM cavities, the relative value of Nb in comparison to the ideal is lower. This outcome is anticipated, as the Nb QWRs were printed using the EB-PBF process, which employs coarser powder, thicker layers, and larger beam diameters. Besides surface roughness, superconducting resonators are sensitive to other loss channels, such as magnetic flux trapping, two-level losses, contamination, and hydride formation. Determining the exact magnitude of each loss without comprehensive measurements is challenging. However, the AM Nb cavities exhibited a Q of approximately 15% of the machined part, with degradation likely stemming from reduced material purity and increased surface roughness. Interestingly, the etched Ti64 *Cavity 4* demonstrated a higher Q than both AMed Nb cavities, despite its lower superconducting transition temperature. Additionally, the Ti64 *Cavity 3*, which underwent extensive abrasive finishing, exhibited the poorest performance. The current hypothesis attributes these increased losses to lossy ceramic media embedded in the RF surface, necessitating further destructive analysis for confirmation. It is important to note that normal conducting measurements, while useful for characterizing surface finish and geometric accuracy, do not directly translate to superconducting performance.

In Ref.[176], researchers 3D printed other cavities using a selective laser melting (SLM) process. The powder composition was 12.18% Si, 0.118% Fe, and 0.003% Cu, with the balance consisting of Al. A high-purity argon gas atmosphere was employed during processing to minimize oxidation. These cavities underwent additional processing. Intriguingly, it was discovered that maintaining rough cavity walls, as printed by the SLM machine, posed no disadvantage compared to smoothing the internal surfaces and polishing with diamond paste. Any observed difference resulted in an improvement of approximately 10%, which falls within the Q-factor measurement error.

The accelerated designs currently under development show immense potential for quantum sensors across various applications[41], yet their scalability to multi-qubit systems, such as quantum computers, remains a challenge. Despite this, the remarkable quality factor observed in 3D niobium MW cavities presents a promising route towards achieving extended coherence in cavity quantum electrodynamics (cQED) architectures for quantum computing. This could be accomplished through proper coupling and demonstration of qubit operations with MW cavities. Considering surface passivation using alternative materials like NbN, NbTiN, and $Nb_3Sn$ may prove beneficial, as these superconductors lack the complex near-surface oxides that seemingly restrict the ultimate high-quality factor in niobium-based cavities. Investigating cavity configurations capable of simultaneous operation in multiple modes with independent tuning and field control is also of great importance.

This architectural approach has the potential to enable "stacking" qubits within a smaller physical volume, while maintaining a wide frequency separation and harmonic relation between modes occupying the same space. Furthermore, coupled cavity systems can be operated with multiple resonances (passbands) of the same mode, exhibiting varying phase advances. This approach can lead to the creation of compact multi-mode systems, where frequency separation is significantly reduced yet remains controllable through adjustments in coupling strength.

## 7. Cavities with normal metals

Thermalization of 3D qubits poses significant challenges due to the poor thermal conductivity of superconductors, unlike normal metals such as copper (Cu). Additionally, the superconducting qubits discussed previously cannot be tuned with an external magnetic flux because superconductors expel magnetic fields. One solution to these issues is to employ cavities made from normal metals. This section explores 3D qubits comprised of normal metals or including normal metal components for magnetic flux tunability and thermalization.

Chad Rigetti et al. reported a 3D qubit system based on a single Josephson junction (JJ) transmon within a Cu waveguide cavity, demonstrating improved lifetimes of $T_1 = 70$ μs and $T_2 = 92$ μs [146]. This improvement resulted from three parallel strategies to enhance coherence limits due to cavity photon-induced dephasing: (i) optimizing qubit and cavity parameters to minimize the expected qubit dephasing rate per residual photon in the fundamental cavity mode, (ii) designing the cavity to reduce the influence of higher modes, and (iii) utilizing Cu material to allow thermalization of the interior cavity walls to very low temperatures of the bulk Cu due to its high thermal conductivity. As the walls are the primary source of dissipation, this reduces thermal noise. The setup's transmission line modes must also be thermalized to sufficiently low temperatures.

In Ref. [177], a 3D rectangular Cu cavity encapsulating a transmon was employed to demonstrate the Autler-Townes effect, also known as the AC Stark effect. The formation of the dark-dressed state with zero eigenvalues was shown with high fidelity. The dark-dressed state is highly intriguing for various quantum applications, including quantum sensing[75] and quantum isolation[178].

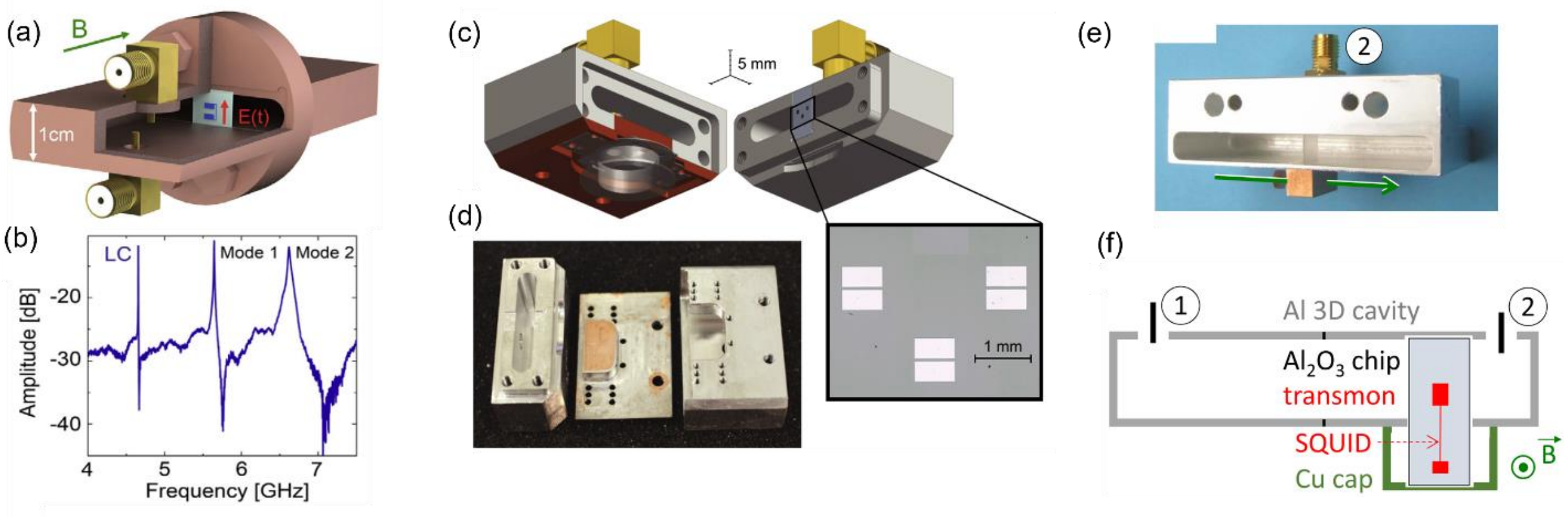

**Figure 3**. **Cavities comprising normal metals.** (a) Cutaway representation of the 3D cavity, with the LC circuit (in blue) on its sapphire chip. The green arrow represents the applied magnetic field B. The red arrow represents the ac electric field E(t) of the first mode of the cavity. (b) Transmission spectrum of the cavity coupled to the LC resonator (no magnetic bias). The first peak corresponds to the LC resonator, while the two other peaks correspond to the first cavity modes[179]. (c) 3D cavity model with a chip containing three flux tunable transmon qubits (inset). The cavity body primarily comprises Al (gray) and features a highly essential Cu insert (brown). The ingenious inclusion of the Cu insert facilitates the efficient penetration of the magnetic field emanating from the three mounted coils, thereby enabling individual control of all qubit frequencies within the cavity's interior. Furthermore, the Cu insert serves the crucial purpose of establishing a reliable thermal link to the chip. (d) Photograph of the cavity parts[85]. (e) Superconducting Al 3D cavity with encapsulated transmon and a part of the SQUID outside the cavity but inside the Cu cap enabling tunability with B-field (f)[180].

Beyond the 3D transmon design, a flux qubit (FQ) in a 3D resonator was reported[179]. FQs attract interest among all superconducting circuits because they provide the largest coupling constants in *hybrid systems* involving electronic spins in semiconductors due to their large magnetic dipole[181–183]. These semiconductors typically consist of nitrogen-vacancy centers in diamond or phosphorus donors in silicon, with quantum coherence lifetimes up to seconds[184,185], or even minutes[186]. The realized Rabi frequency in such hybrid systems is considerably larger than the spin dephasing rates. This area requires FQs with $T_2 \geq 5$ μs, which is challenging due to the intrinsically rapid dephasing of FQs. In Ref.[179], a rectangular 3D cavity made of non-superconducting Cu to enable the application of an external magnetic field B to the FQs was reported**, Fig. 3a**. The cavity was coupled to a superconducting aluminum (Al) loop intersected by four Josephson junctions. The transmission spectrum of the cavity coupled to the LC resonator without magnetic bias is presented in **Fig. 3b**. The first peak at frequency corresponds to the resonance of the LC resonator, while the other two peaks relate to the first modes of the cavity. The measured qubits reached reproducible coherence times $T_2 = 2...8$ μs, sufficient to achieve the strong-coupling regime with a single spin. The hybrid device incorporates a comparable design, featuring a superconducting transmon qubit and a mechanical resonator, coupled through the utilization of magnetic flux[187].

A similar approach, employing a 3D rectangular Cu resonator with Q of $6\times10^3$ (undercoupled) and coupled to a C-shunt flux qubit with substantially reduced charge noise, was reported[60]. This approach demonstrated relatively high quantum coherence times of 80 μs (Hahn echo). The qubit energy relaxation was limited by quasiparticle tunneling and fluctuating TLSs, while its dephasing resulted from charge noise and critical-current fluctuations. The obtained value of the Purcell-limited relaxation time of 90 μs indicates that the system can be improved if all other damping mechanisms are mitigated.

A hybrid approach to superconducting cavities with parts made of normal metals has garnered increasing attention, as it allows for the preservation of the high Q-factor of superconducting cavities while benefiting from the magnetic field tunability and excellent thermalization characteristics of normal metals. **Fig. 3c** illustrates such a design of a 3D cavity with a chip containing three flux-tunable transmon qubits (inset)[85]. Most of the cavity is made of Al (gray), with a Cu insert (brown). **Fig. 3d** presents a photograph of the cavity parts. The Cu insert enables the penetration of the B-field from the three mounted coils into the cavity's interior, allowing precise control of each qubit's frequencies. Moreover, it establishes a robust thermal connection to the chip, ensuring efficient heat dissipation.

The tunability of a superconducting qubit within an entirely superconducting cavity was successfully achieved[180], **Fig. 2e**. This was accomplished by placing a part of the chip with the SQUID loop outside of the cavity but inside the Cu cap, enabling tunability with the B-field, as shown in **Fig. 2f**. Although this approach allows for the preservation of the cavity's high-Q features, its practicality is limited by the requirement for a qubit design tailored to the method, which may hinder scalability to large-scale, many-qubit systems.

To improve the noise characteristics of 3D cavity qubits, the suppression of coupling with input and output ports using attenuators has been actively employed. Such attenuators dissipate excess photons in the readout mode and suppress photon-induced qubit dephasing. Resonant attenuators can also function as Purcell filters, filtering off non-resonant photons[104]. Interestingly, an attenuator with excellent thermal conductivity can also be implemented in the 3D design and supplied to the 3D qubit. In Ref.[97], a band-pass

MW attenuator consisting of a dissipative cavity well thermalized to the mixing chamber stage of a dilution refrigerator was designed and realized. The high coherence lifetimes of 230 μs (Hahn echo) and 40 μs (Ramsey) were reported at a temperature of approximately 20 mK for a modest Q-factor.

## 8. Multiple 3D-qubits

Although several important advances have recently been demonstrated using 3D qubits, including extending the lifetime of an error-corrected qubit beyond its constituent parts[18], randomized benchmarking of logical operations[188], a CNOT gate between two logical qubits[189], and Ramsey interference of an encoded quantum error-corrected qubit[190], scaling to systems with more coupled qubits remains much less explored. This section reviews the recent results in 3D multi-qubit and multi-state systems.

A robust two-cavity design has been developed[191], in which the superconducting transmon qubit is inserted into both cavities, as depicted in **Fig. 4a**. The system comprises two 3D rectangular Al (alloy 6061) waveguide cavities coupled to a vertical transmon qubit, which was fabricated on a c-plane sapphire substrate. The qubit's coupling strength was determined by the length of the stripline antenna, which extends into each cavity. This configuration exhibits remarkable Kerr nonlinearity due to the qubit anharmonicity and the consequent nontrivial coherent dynamics, culminating in the observation of the collapse and revival of a coherent state. The origin of this phenomenon can be attributed to two fundamental factors: the quantization of the light field confined within the cavity, and the intricate interplay of nonlinear interactions between individual photons. One cavity functions as a storage cavity (quantum memory), and the other serves as a readout cavity. Both cavities are carved out of Al (alloy 6061 T6). The cavities have a total quality factor of $Q_i \approx 1\times10^6$, limited by internal losses.

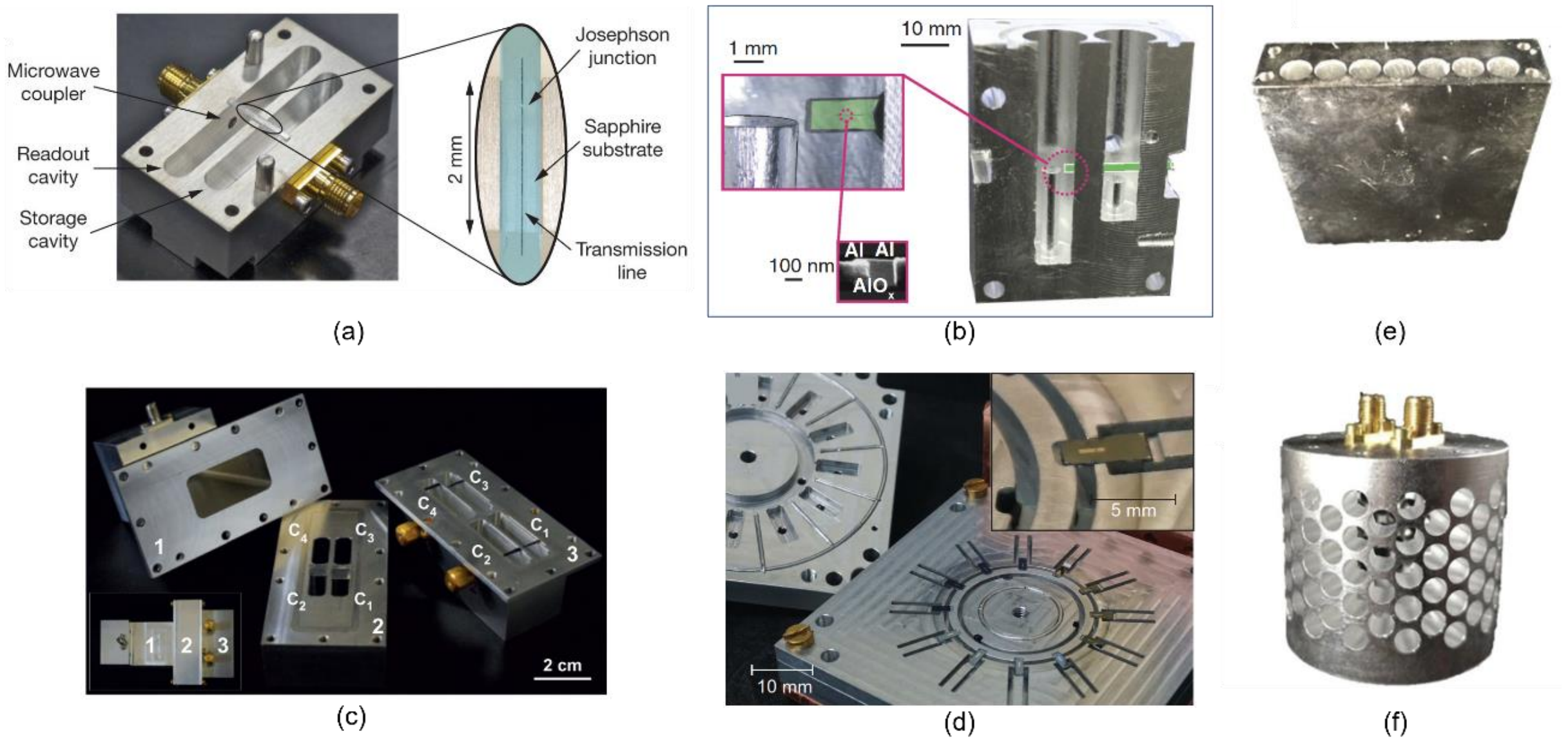

**Figure 4. Multi 3D qubit systems**. (a) Two Al waveguide cavities made of alloy 6061 are coupled to a vertical transmon qubit, with one half of each cavity connected to the qubit. A magnified view of the specified area in the photograph reveals a closer look at the qubit, which was fabricated on a c-plane sapphire substrate measuring 1.4 mm in width, 15 mm in length, and 430 μm in thickness. The coupling strength of the qubit is determined by the length of the stripline coupling antenna, which extends into both cavities. (b) Cross-section of the system of two cavities integrated with a transmon (green), with insets that show the transmon chip and a micrograph of the junction[30]. (c) Four-cavity structure made of Al in three separate pieces. Piece 1 is the rectangular waveguide. Piece 2 (3) comprises one (second) half of the readout cavities ($C_{1–4}$). The qubits on Si chips were positioned to achieve a similar dispersive shift for each cavity. Inset: fully assembled structure[192]. (d) Semi-assembled 3D cQED implementation of a 12-qubit network[193]. (e), (f) Pictures of flute monolithic superconducting MW cavities of rectangular (e) and cylindrical (f) geometry[194].

In another work[30], a system of two coupled 3D cavities in the form of cylindrical coaxial cavities made of pure Al interacting with a joint qubit (transmon) has been experimentally investigated, as shown in **Fig. 4b**. The resonators have cylindrical geometry and provide a high Q-factor up to $7\times10^7$ (single-photon limit) and $2\times10^8$ (many photons). The saturation of the $Q_i$ indicates that the TLSs are the primary source of loss in this system. The authors conclude that this system shows no additional, unknown dephasing channels arising from connecting the individual constituents and that the coherence properties of the system can be improved by employing better transmons or cavities. A similar coaxial geometry of a single 3D cavity made from 5N Al but with a half-sphere on top of the post is shown to reduce the amplitude of the E-field on the surface of the post and slightly increase the intrinsic Q-factor[31].

In Ref.[192], the simultaneous measurement of four transmon qubits in four individual cavities coupled to a single rectangular waveguide has been reported, as shown in **Fig. 4c**. The measured values of the lifetimes were approximately 3 μs in Hahn echo and ~2.5 μs in the Ramsey experiment. Despite these relatively small values and modest Q-factor of the cavities (~1,000), this is an important demonstration of the coupling of four 3D qubits, potentially enabling a plethora of nontrivial multi-qubit quantum effects.

Exploring ring resonators provides a route to coupling more qubits in a single device. In Ref.[193] such a 3D ring resonator coupled with 12 qubits has been realized, as shown in **Fig. 4d**. The transmon qubits were placed $30^{\circ}$ apart, and the readout resonators were $\lambda/4$ sections of the transmission line extending radially outward. The four parts of the resonator were machined from pure Al, and the transmon qubits demonstrated relatively high coherence characteristics [~3.5 μs (Hahn echo) and ~29 μs (Ramsey)] for such a complex 12-qubit architecture. Such a system of 12 fully controlled and coupled qubits is very promising for highly connected qubit networks for superconducting circuits.

The extension of cQED to many cavity modes (multimode cQED) promises explorations of many-body physics with exquisite single-photon control[194]. Multimode cQED systems with strong light-matter interactions have been realized in various 2D quantum circuits, with a Josephson-junction-based superconducting qubit coupled to many nearly harmonic modes, especially in quantum metamaterials[195–197]. In Ref.[194], a new approach of flute geometry to the seamless high-Q MW cavities for multimode circuit quantum electrodynamics has been proposed. Cavities of rectangular (**Fig. 4e**) and cylindrical (**Fig. 4f**) geometry have been realized and coupled to a single transmon qubit. Coherent coupling of the qubit to 9 cavity modes was demonstrated. Due to the seamless realization and high purity of Al, the system possesses high-Q in the order of $10^8$ and a long quantum coherence lifetime (~2-3 ms, Ramsey). These results are nearly two orders of magnitude better than those reported in multimode cQED systems. The coherence lifetimes of these cavity modes are comparable to the longest reported in single or few-mode 3D cQED systems.

## 9. Bosonic-Encoded Qubits

3D cavities with high-Q modes have recently gained significant importance for the implementation of BECs in quantum computing and communication systems. BEC, a unique type of quantum bit, leverages the properties of bosons, a particle class that adheres to Bose-Einstein statistics. In contrast to conventional qubits based on two-level systems, such as electron spin states, BECs are encoded in multiphoton states of QHOs. These oscillators can include electromagnetic fields in resonators or mechanical vibrations in solid-state devices.

A primary advantage of BEC lies in their ability to utilize the symmetries of multiphoton states to protect the encoded quantum information from noise and decoherence. By carefully choosing multiphoton states with specific symmetries, these qubits can be made highly resistant to certain error types[198]. This error-resilient characteristic is extremely desirable for quantum computing and communication applications, as it significantly reduces the need for complex error-correction schemes.

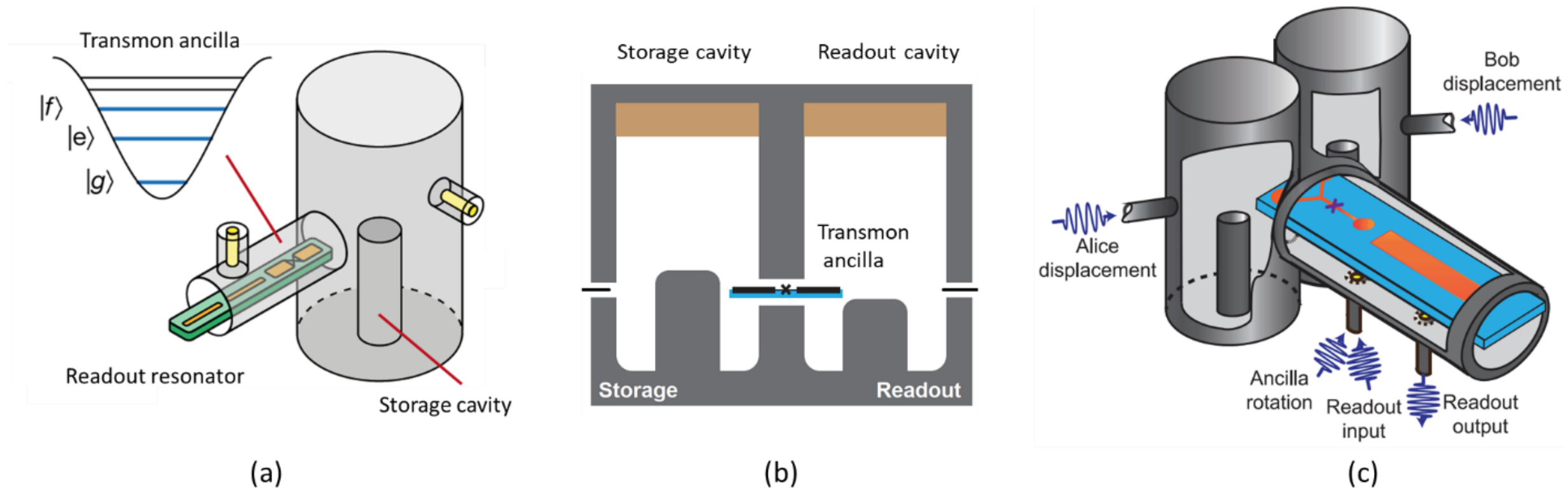

**Figure 5**. **3D cavities for bosonic-encoded qubits**. (a) 3D cavity combined with a transmon sapphire chip, readout resonator, and Purcell filter, is utilized in this setup. The cavity's electromagnetic mode implements a QHO, while the transmon's $|g\rangle$ and $|e\rangle$ levels function as an auxiliary qubit, assisting in oscillator quantum error correction (QEC)[199] (b) Two coaxial MW resonators (gray) bridged by a single transmon superconducting qubit (black) fabricated on a sapphire chip are anchored on a dilution refrigerator base plate[200]. (c) 3D view of the device consisting of two coaxial cavities (Alice and Bob), a Y-shaped transmon with a single Josephson junction (marked by ×), and a stripline readout resonator[201].

BEC, which are encoded in the infinite-dimensional Hilbert space of a QHO, present a promising approach for quantum information processing[34,200,202]. In this encoding, superpositions of multiphoton states in the QHO, with modes following bosonic statistics, define the qubit. The QHO is typically implemented using an engineered electromagnetic mode in a 3D MW cavity or a lithographically-defined transmission line resonator on a 2D chip. Although QHO qubits lack individually addressable energy-level transitions, complicating qubit manipulation compared to transmons, the QHO qubit is coupled to a 2D transmon ('ancilla') that enables control and readout. This combination allows for universal control through MW irradiation and manipulations of the coupled transmon. Additionally, the extended lifetimes of MW cavities make this encoding highly attractive for quantum information processing[30].

These cavities possess a large Hilbert space for encoding information in multi-photon states compactly, forming a logical qubit within a single piece of hardware. The utilized 3D cavity can technically be any of the resonators discussed in the previous section. The resonator is coupled to external excitations through an anharmonic controlling element. The anharmonic element, typically in the form of a transmon and henceforth referred to as the '*ancilla'*, provides the necessary non-linearity for controlling and measuring cavity states[199,203,204], **Fig. 5a**. In this case, the artificial atoms function as non-linear ancillae, enabling conditional operations and efficient tomography of the cavity states. The anharmonic element can also be a 3D transmon, with the entire structure realized as a cavity featuring two chambers connected via a 2D transmon[18,190], like the one shown in **Figs. 4a,b**. This approach is illustrated in **Fig. 5b**. Of particular interest is the case of coupling multiple 3D resonators through one or several ancilla qubits. This geometry allows for boson encoding in so-called cat states. A key motivation for creating multicavity cat states is to implement a promising paradigm for fault-tolerant quantum computation, wherein information is redundantly encoded in the quasi-orthogonal coherent state basis[205]. This approach has recently led to the first realization of quantum error correction for a logical qubit, achieving the break-even point at which the qubit's lifetime surpasses the lifetime of the system's constituents[18].

## 10. Overlook and perspectives

In this comprehensive study, we examine the advancements in the domain of 3D qubits, focusing on high and ultra-high Q-factor systems. We provide an overview of superconducting Josephson junction (JJ) qubits and the progress in 2D topological structures. Additionally, we explore the damping and noise mechanisms in resonators and qubits that contribute to quantum state dephasing. Our investigation encompasses 3D cavities with potential quantum applications, along with existing implementations of single 3D qubits made from superconducting materials and those composed of conventional materials. Furthermore, we delve into the multi-qubit, multi-state quantum systems and BECs developed thus far.

We present a summary of key findings in designing MW 3D cavities for circuit quantum electrodynamics (cQED) applications in **Table 3**.

**Table 3**. Characteristics of the realized 3D cavity for qubits; sp stands for "single photon".

| Resonator | Material/Fabrication | $Q_i$ | Qubit | Max $T_2$ | Year, Ref. |
|---|---|---|---|---|---|
| Rectangular | Al | $5\times10^6$ | Transmon | 20 μs | 2011, [52] |
| Cylindrical | Al 5N5, acid treated | $7\times10^7$ (sp) | No | -- | 2013, [84] |
| Rectangular | Al 5N5, acid treated | $43\times10^6$ (sp) | No | -- | 2013, [84] |
| Rectangular | Cu, oxigen-free | $1.8\times10^4$ | Transmon, 3 levels | 51 μs | 2013, [177] |
| Rectangular (two) | Al (alloy 6061 T6) | $1\times10^6$ | Transmon | 8 μs | 2013, [191] |
| Rectangular | Cu | $1.5\times10^4$ | Flux qubit | 8 μs | 2014, [179] |
| Rectangular | Al (alloy 6061) with a small insert Cu (C10100) | $\sim10^5$ | No | -- | 2016, [85] |
| Coax-line architecture | Al, chemically etched | $11.2\times10^6$ | Transmon | 60 μs | 2016, [206] 2016, [201] |
| Cylindrical coaxial | Al, 4N | $7\times10^7$ (single-photon) $2\times10^8$ (many photons) | BEC (Transmon ancilla) | 720 μs ($T_1$=1.2 ms) | 2016, [30] |
| Rectangular | Cu | $4\times10^7$ | Fluxonium | 4 μs | 2018, [107] |
| Rectangular (with attenuator) | Al (cavity) Brass (attenuator) | -- 500 | Transmon | 230 μs (Hahn echo) 40 μs (Ramsey) | 2019, [97] |
| Rectangular | Cu | $6\times10^3$ (undercoupled) | Flux qubit (C-shunt) | 80 μs (Hahn echo) 18 μs (Ramsey) | 2019, [60] |
| Rectangular (four) | Al | $\sim1\times10^3$ | Transmon (x4) | ~3 μs (Hahn echo) ~2.5 μs (Ramsey) | 2019, [192] |
| TESLA (single-cell) | Nb, fine-grain | $4\times10^{11}$ (sp) | No | -- | 2020, [86] |
| *RadiaBeam* (coaxial quarter-wave cavity with the optimized shape) | Nb | $\sim1\times10^{7}$ | No | -- | 2020, [67] |
| 3D stub-geometry | Al | $11.5\times10^{6}$ (sp) | No | -- | 2020, [31] |
| Coaxial quarter-wave cavity (tunable by sapphire rod) | Al, 99.999%-purity | $1.3\times10^{7}$ | Transmon (integrated with electro-optic transducer) | ~20 µs | 2021, [207] |
| 3D Ring-Resonator | Al | -- | Transmon (x12) | ~3.5 µs (Hahn echo) ~29 µs (Ramsey) | 2021, [193] |
| Multimode flute cavity<br>Rectangular<br>Cylindrical<br>Coaxial | Al<br>5N5<br>6N<br>5N | $6.5-9.5\times10^{7}$<br>$2.54\times10^{7}$<br>$9.79\times10^{7}$ | Transmon | ~2-3 ms (Ramsey) ($T_1$=80–100 µs) | 2021, [194] |
| Rectangular | Cu | 377 | Fluxonium | ~1.48 ms | 2021, [33] |
| Quarter wave resonator (QWR), *RadiaBeam* | Nb Machined<br>Nb EB-PBF<br>Ti64 L-PBF | $9.2\times10^{6}$<br>$1.6\times10^{6}$<br>$2.2\times10^{6}$ | No | -- | 2022,[174] |

Upon evaluating this data, it becomes evident that significant progress has been made—from modest Q cavities yielding relatively brief quantum coherence lifetimes to ultra-high Q-factor cavities boasting prolonged qubit coherence. Based on our analysis, we draw the following conclusions:

1. Although cavities composed of or containing normal metals with intrinsically superior thermal conductivity offer enhanced thermalization and reduced effective qubit temperatures, superconducting cavities remain the preferred choice.
2. Investigating novel materials, such as tantalum (Ta), can substantially improve the lifetime characteristics of 3D transmons, similar to their 2D counterparts[94].
3. Cylindrical or coaxial seamless 3D cavities exhibit superior characteristics due to their reduced surface participation factor and absence of edges.
4. The performance of cylindrical cavities can be further optimized through cavity shape design.
5. While the 3D cavity approach suffers from limited scalability, it holds significant potential for quantum memory and sensor applications. Recent developments in the multilayer MW integrated quantum circuit approach show immense promise for multi-qubit 3D architectures[208].
6. Several promising 3D cavity designs have yet to be integrated with qubits, suggesting substantial future progress in this field.

Upon comprehensive examination of and conclusions, it is clear that substantial research endeavors are required to augment the attributes of JJ and BEC systems. Key avenues for advancing the field could include: (i) probing innovative materials for cavities and qubits, (ii) devising scalable fabrication methodologies, (iii) discovering novel resonant modes and modes exhibiting nontrivial topology, (iv) exploring new EM scattering effects for excitation and control. In light of this, here we succinctly discuss the perspective of these approaches.

In the pursuit of novel materials, a range of potential superconductors has been studied for constructing JJ cQED circuits, including tantalum[94], as well as compounds like TiN[149], NbN[209], NbTiN[210], and granular Al[211]. Recent research showcased the considerable potential of tantalum (Ta) [212] by fabricating a 2D transmon qubit with an exceptional $T_1$ lifetime of ~500 μs and a qubit dephasing time $T_2$ of 557 μs. Nevertheless, the observed lifetimes differ considerably between devices, even on the same chip, owing to fluctuations in the fabrication process. Comparisons with alternative materials, such as Nb and Al, using identical design and fabrication procedures, reveal that qubits produced with Ta films exhibit superior performance.

As quantum circuits grow in complexity, the need for precise control over couplings, reduced cross-talk, and minimized coupling to undesirable modes becomes increasingly vital. Attaining the scalability of electronic components, akin to the capabilities of CMOS technology, presents a comparable challenge in the context of superconducting cQED devices. Effectively coupling multiple circuit elements through manageable channels while concurrently suppressing any unintended interactions remains a daunting task. Nevertheless, the enclosed 3D waveguide packaging for cQED devices has demonstrated immense potential in tackling these hurdles and facilitating the development of sophisticated quantum systems with high coherence[206]. By incorporating planar circuit elements within a 3D waveguide package, the resulting coaxial transmission line-based device can capitalize on the intricacy and dimensional control of planar systems while also enjoying the coherence, coupling, and spectral purity of 3D systems. Furthermore, this implementation experiences minimal impact from mechanical assembly uncertainties. This system has already been successfully adapted for more elaborate experiments involving multiple qubits or cavities[201]. The fabrication of resonators on c-plane sapphire chips using photolithography, evaporated Al 80 nm thick deposition, and a lift-off process, accompanied by the machining of the enclosure from high-purity Al and chemical etching, exemplifies the viability of this approach. As quantum circuits become increasingly intricate, the adoption of the 3D waveguide package architecture promises to be an invaluable asset in achieving the requisite level of control and coherence.

While 3D resonators and BECs provide extended coherence times, 2D planar resonators are better suited to existing microfabrication processes. Research on multilayer structures aims to integrate both approaches [213]. One potential method involves employing Si wafer micromachining to embed quantum circuit components within a multi-wafer construction containing numerous JJ qubits, memories, buses, and amplifiers. Refs.[208,214] propose the multilayer MW integrated quantum circuit (MMIQC) hardware platform for scalable micromachined enclosures. In MMIQC, 3D resonators are micromachined within wafers of varying planes, creating vacuum gaps for energy storage. This technique merges the benefits of integrated circuit fabrication and the lengthy coherence times attainable in 3D cQED.

We would like to highlight two recent findings that hold great promise for cQED systems in exploring new types of resonant modes and modes with nontrivial topology. First, magnetic localized spoof plasmon (LSP) skyrmions have been identified in space-coiling meta-structures[215]. These skyrmions arise from the tailored metastructure's eigen-resonances and display multiple-π-twist vectorial configurations in real space, yielding subwavelength features down to $\lambda^3/10^6$ via near-field scanning. Notably, LSP skyrmions possess immense topology stability and exhibit a stable multi-resonant spectrum with adaptable skyrmionic textures. Additionally, these 2D skyrmionic meta-resonators support ultra-high Q-factor modes, offering significant potential for improvement in the superconducting regime. Impressively, 2D coplanar meta-structures demonstrate a Q-factor comparable to the best 3D resonators for qubits, and even surpass them in tunability and scalability when integrated with a JJ qubit[215].

The second phenomenon, known as embedded eigenstates or bound states in the continuum (BICs), has garnered considerable attention. BICs, predicted by von Neumann and Wigner, are an eigen-solution of the single-particle Schrödinger equation that exists within the continuum despite being compatible with decay in terms of momentum[216]. BICs have been experimentally observed in various wave settings due to mathematical analogies with the wave equation in electrodynamics, optics, and linearized acoustics[217]. BICs represent a resonant state of an open system supporting an infinite radiative Q-factor while maintaining momentum compatibility with radiation[218]. Symmetry-protected BICs can form at the Γ-point of a periodic structure[219]. A periodic array of resonators, such as metallic cylindrical disks, can create a symmetry-protected BIC by directing scattered power towards a single direction due to destructive interference among array elements. The resonant mode sustains vertical dipoles with the same amplitude and phase, rendering them non-radiating. Small broken symmetries along the array can trigger coupling to outgoing radiation, resulting in a sharp Fano line shape in the radiation spectrum, often referred to as a quasi-BIC[220–222]. Given the theoretically unbounded radiative Q-factor values of BIC modes, their coupling with JJ qubits offers exciting potential for novel qubits.

Concerning readout mechanisms, current quantum computers heavily rely on nonreciprocal devices, isolators, and circulators to shield the ~10 – 100 mK quantum processor from room temperature noise. Presently, nonreciprocal components primarily employ the Faraday effect, utilizing ferrite materials with static magnetic field bias[223]. However, these devices are lossy, bulky, minimally tunable, and incompatible with planar technologies, including transmission-line quantum circuits. Furthermore, magnetic isolators and circulators introduce challenges in superconducting quantum circuits due to loss (noise) and the circuits' sensitivity to static magnetic fields. Thus, there is a pressing need for new materials and devices enabling magnet-less nonreciprocal responses in quantum systems[178].

Another promising research avenue involves developing efficient electro-optic transducers for optical readout of JJ qubits' states. Optical quantum networks excel at transmitting quantum states over long distances without requiring low temperatures[224–226]. Consequently, a quantum-enabled electro-optic transducer bridging the MW and optical domains would significantly enhance quantum devices' capabilities. Electro-optic elements have shown promise for delivering classical signals to superconducting circuits[116,227]. Devices for transducing quantum states to optical signals have also been proposed[207,228]. Notably, mechanical resonators can serve as transducers in hybrid setups between various quantum systems[229] and for converting between MWs and optical light[230–232]. Recent work[207] reported an electro-optomechanical transducer embedded in a 3D cQED architecture on a single chip. The cQED system features a planar superconducting transmon qubit coupled to a seamless 3D Al cavity's fundamental mode. Impressively, tunability in this 3D qubit system is achieved by translating a sapphire rod within the 3D cavity using a piezoelectric stepping module. The electro-optic transducer comprises a MW resonator, an optical resonator, and a single mode of a micromechanical oscillator.

Intense research efforts are focused on devising methods to render superconducting 3D cQED architectures more controllable and tunable. The magnetic hose concept, for instance, has been proposed for rapid magnetic flux control within a superconducting waveguide cavity[233], designed as an effective MW filter to preserve the qubit's energy relaxation time. This approach enables control of 3D transmon on time scales under 100 ns while maintaining a high Q-factor. Other techniques for 3D cavity adjustment include coupling pins[137] and sapphire rods manipulated by piezoelectric stepping modules[207].

The unique geometry of 3D cavities enables the exploration of material characteristics at ultra-low temperatures. Recent studies have investigated polariton states and strong light-matter interactions within such cavities[234]. Coupling different materials, such as c-Si, with 3D cavities may enable the utilization of Mie resonances, which offer low-loss, dynamic magnetic response, and a rich spectrum of available modes[235–239].

Longer coherence times can be achieved by coupling JJ qubits with highly coherent systems to create hybrid quantum systems[44]. *Quantum phononics*, which involves coupling distant quantum systems coherently using phonons, is another area of interest. Coherent generation and measurement of complex stationary phonon states have been demonstrated with JJ qubits[240,241].

Finally, proficient control over excitations in QHO modes is crucial for advancing quantum systems that employ bosonic codes and encoded qubits. cQED architecture encodes data in multi-photon states within superconducting MW cavities, necessitating meticulous control of each mode. As such, it is essential to examine the impacts of virtual perfect absorption and virtual critical coupling, recently proposed to bolster these quantum systems. Coherent perfect absorption has enabled efficient wave absorption control by exciting a real-frequency-axis scattering matrix zero[218,242]. Adjusting the coherence and relative phase of counterpropagating beams exiting the absorber allows for the implementation of efficient optical switches and logic gates. Notably, recent research has shown the feasibility of inducing a similar response in lossless systems by engaging scattering zeros in the complex frequency plane. Tailoring the incident field in time to match the exponentially growing fields associated with a complex zero results in a transient where the structure's scattering vanishes, as though it were perfectly absorbing, despite lacking loss or absorption. The system stores incoming electromagnetic energy instead of converting it to other forms. Altering the excitation or the impinging waves' relative phase releases the stored energy on demand. This approach, known as coherent virtual absorption (CVA) [218,243], is the sole method for achieving *unit excitation efficiency* in high-Q resonators[244,245]. CVA can be employed for efficient light storage and release in quantum bosonic memories and BEC. Consequently, these CVA breakthroughs present promising prospects for enhancing quantum systems and BECs, underscoring the significance of exploring innovative effects and pushing the limits of quantum system capabilities.

## Acknowledgment

The work on this paper was supported by the U.S. Department of Energy, Office of High Energy Physics, under SBIR grants DE-SC0018753 and DE-SC0019973. The work at Jefferson lab is supported by Jefferson Science Associates, LLC under U.S. DOE Contract No. DE-AC05- 06OR23177. AF was supported by , the ARC Centre of Excellence for Engineered Quantum Systems (EQUS, No. 286 CE170100009) and by , and the Foundational Questions Institute Fund (Grant No. FQXi-IAF19-04).

### AUTHOR DECLARATIONS

### Conflict of Interest

The authors have no conflicts to disclose.

### DATA AVAILABILITY

Data sharing is not applicable to this article as no new data were created or analyzed in this study.